\documentclass[a4paper]{article}

\usepackage[english]{babel}
\usepackage{geometry}  
\usepackage{graphicx}      
\usepackage{tabularx}      
\usepackage{amssymb}
\usepackage{amsmath}
\usepackage{xltabular}
\usepackage[hyphens]{url}
\usepackage[backend=biber,style=nature, doi=false,isbn=false, date=year, url=false, autopunct=true]{biblatex}
\makeatletter
\renewbibmacro*{textcite}{%
  \iffieldequals{namehash}{\cbx@lasthash}
    {\usebibmacro{cite:comp}}
    {\usebibmacro{cite:dump}%
     \ifbool{cbx:parens}
       {\bibclosebracket\global\boolfalse{cbx:parens}}
       {}%
     \iffirstcitekey
       {}
       {\textcitedelim}%
     \usebibmacro{cite:init}%
     \ifnameundef{labelname}
       {\printfield[citetitle]{labeltitle}}
       {\printnames{labelname}}%
     \addspace
     \ifnumequal{\value{citecount}}{1}
       {\usebibmacro{prenote}}
       {}%
     \mkbibsuperscript{\usebibmacro{cite:comp}}%
     \stepcounter{textcitecount}%
     \savefield{namehash}{\cbx@lasthash}}}
\makeatother

\usepackage{csquotes}
\usepackage{siunitx}
\usepackage{geometry}
\usepackage{multirow}
\usepackage[dvipsnames,table,xcdraw]{xcolor}

\usepackage{xspace}
\usepackage{caption}
\usepackage{subcaption}
\usepackage{authblk}

\usepackage{marvosym}

\usepackage{sectsty} 

\sectionfont{%
  \fontfamily{phv}\selectfont
  \Large
  \bfseries             
}

\subsectionfont{%
  \fontfamily{phv}\selectfont%
  \large
  \bfseries
}

\definecolor{halfgray}{gray}{0.55}

\IfFileExists{./fonts/Heliotrope-Bold.otf}{
    \usepackage{fontspec}
    \setmainfont{Heliotrope}[Extension = .otf, UprightFont = *-Regular, ItalicFont = *-Italic, BoldFont=*-Bold, Path = ./fonts/]
}{
    \usepackage{biolinum}
    
}

\usepackage[hyphens]{url}
\usepackage{xcolor}

\definecolor{linkcolor}{RGB}{46, 72, 224}

\usepackage[colorlinks=true,
  linkcolor=linkcolor,
  citecolor=linkcolor,
  urlcolor=linkcolor]{hyperref}

\usepackage[nameinlink,noabbrev]{cleveref}

\newcounter{SIsec}
\renewcommand{\theSIsec}{S\arabic{SIsec}}

\crefname{SIsec}{Section}{Sections}
\Crefname{SIsec}{Section}{Sections}

\newcommand{\SIsection}[2]{%
  \clearpage
  \refstepcounter{SIsec}%
  \subsection*{\theSIsec.\ #1}%
  \phantomsection%
  \label{#2}%
}

\usepackage{lipsum}
\usepackage[labelfont=bf]{caption}

 \usepackage{microtype}
 \usepackage{fontawesome}

\definecolor{linkcolor}{RGB}{46, 72, 224} 

\usepackage[version=3]{mhchem} 
\usepackage{graphicx}
\usepackage{orcidlink}
\usepackage{multirow}
\usepackage{threeparttable}
\usepackage{arydshln}
\usepackage{xcolor}
\usepackage{booktabs}
\usepackage{physics}
\usepackage{braket}
\usepackage{siunitx}
\usepackage{float}
\usepackage{multicol} 
\usepackage{geometry} 
\usepackage{stfloats}

\title{{\fontfamily{phv}\selectfont\bfseries ML-guided screening of chalcogenide perovskites as solar energy materials}}

\author[1,2,\Letter]{Diego~A.~Garzón\orcidlink{0000-0001-9576-8183}}
\author[3]{Lauri~Himanen\orcidlink{0000-0002-3130-8193}}
\author[2]{Luisa~Andrade\orcidlink{0000-0001-5750-1285}}
\author[1]{Sascha~Sadewasser\orcidlink{0000-0001-8384-6025}}
\author[3]{José~A.~Márquez\orcidlink{0000-0002-8173-2566}}

\affil[1]{International Iberian Nanotechnology Laboratory, Av. Mte. José Veiga s/n, 4715-330 Braga, Portugal}
\affil[2]{LEPABE, ALiCE, Faculty of Engineering, University of Porto, Rua Dr. Roberto Frias, 4200-465 Porto, Portugal}
\affil[3]{Physics Department and CSMB, Humboldt-Universität zu Berlin, Zum Großen Windkanal 2, D-12489 Berlin, Germany}

\affil[\Letter]{\texttt{diego.garzon@inl.int}}

\begin{document}
\maketitle

\begin{abstract}
Chalcogenide perovskites have emerged as promising absorber materials for next-generation photovoltaic devices, yet their experimental realization remains limited by competing phases, structural polymorphism, and synthetic challenges. Here, we present a fully data-driven and experimentally grounded screening and ranking framework to assess the stability and experimental feasibility of chalcogenide perovskites, integrating interpretable analytical descriptors, machine-learning models, and sustainability metrics. Using a curated experimental dataset of halide and chalcogenide compounds, we derive a new tolerance factor via the SISSO (sure independence screening and sparsifying operator) algorithm that more accurately distinguishes perovskite-forming compositions than established tolerance-factor-based screening criteria. This descriptor is combined with generative crystal structure prediction, composition-based bandgap estimation, and machine-learning-based feasibility assessment to systematically explore a wide chemical space of hypothetical chalcogenide perovskites. The resulting candidates are further evaluated using sustainability indicators, enabling multi-objective ranking tailored to both single-junction and tandem photovoltaic architectures. Beyond identifying several promising and previously unexplored chalcogenide perovskites, this work demonstrates a transferable screening strategy for chemically constrained materials spaces that balances optoelectronic performance, experimental viability, and long-term sustainability.
\end{abstract}

\begin{multicols}{2}

\section{Introduction}

\begin{figure*}[ht]
  \centering
  \includegraphics[width=\textwidth]{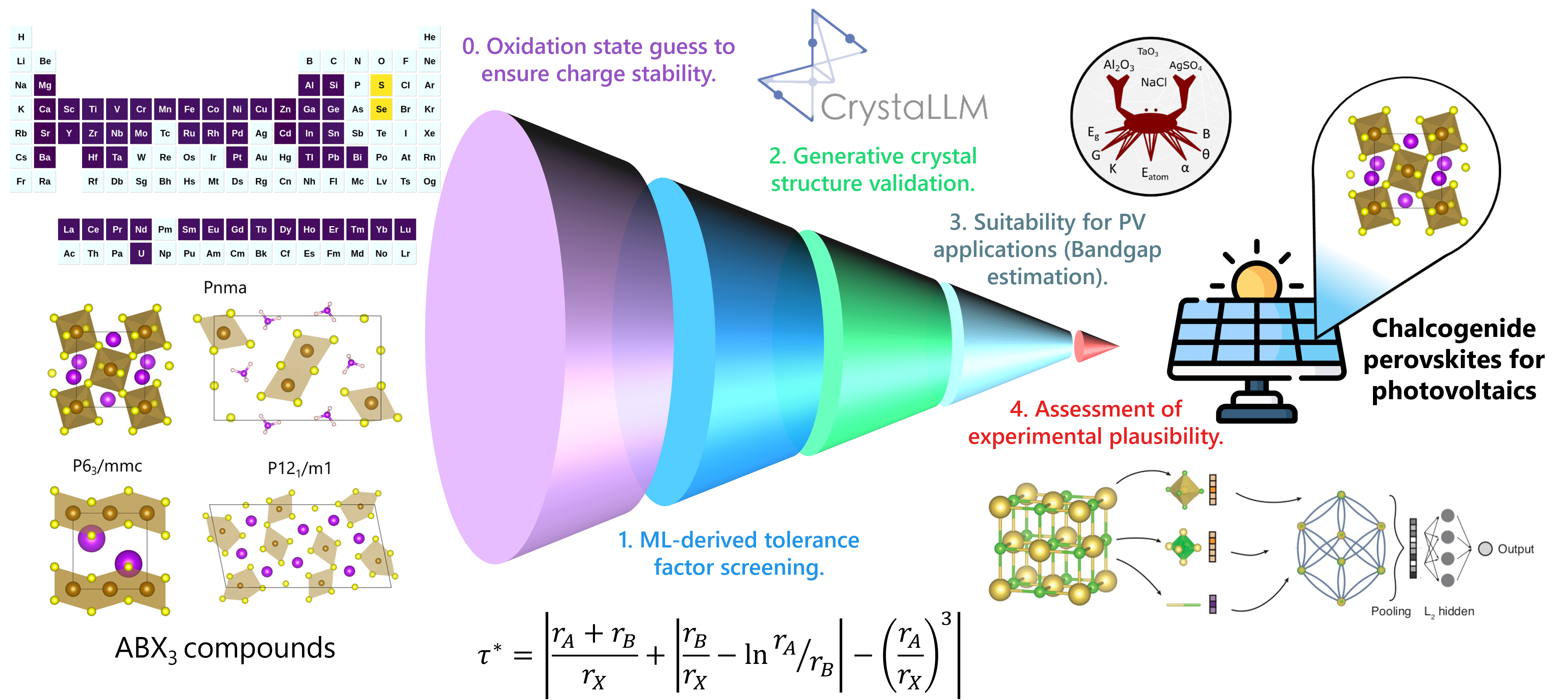}
  \caption{Overview of the ML-guided screening pipeline employed to assess chalcogenide perovskites, combining experimentally motivated descriptors and machine-learning models to evaluate structural stability, crystal structure, experimental plausibility, and photovoltaic suitability.}
  \label{fig:workflow}
\end{figure*}

Chalcogenide perovskites have recently emerged as a promising class of inorganic materials for optoelectronic and energy conversion applications, particularly photovoltaics. In these compounds, the anions occupying the X-site in the ABX$_3$ perovskite-type structure are chalcogenides (X = S$^{2-}$, Se$^{2-}$, Te$^{2-}$), excluding oxygen-based systems. Despite less than a decade of experimental development, chalcogenide perovskites have attracted significant attention due to their intrinsic structural robustness, tunable optoelectronic properties, and absence of toxic elements~\cite{Sopiha_2022}.  

In the context of thin film solar cell technology, hybrid inorganic-organic halide perovskites currently dominate the research field, having achieved a record laboratory efficiency of 27\%~\cite{green2025}. However, concerns regarding their long-term stability, ion migration, and toxicity primarily due to the presence of lead, continue to hinder large-scale deployment~\cite{chowdhury2023}. These limitations have motivated the search for alternative perovskite-inspired absorbers that combine high performance with enhanced chemical and thermal stability. In this regard, chalcogenide perovskites emerge as a compelling candidate, although whether they can ultimately deliver comparable photovoltaic performance remains an open question.

Among chalcogenide perovskites, BaZrS$_3$ is the most extensively studied compound, with a reported bandgap in the range of 1.7–1.95~eV. BaZrS$_3$ exhibits a high optical absorption coefficient above the bandgap, excellent stability under exposure to air, moisture, and heat, and is composed of earth-abundant, non-toxic elements \cite{agarwal2025_bazrs3_review}. These characteristics make it particularly attractive as an absorber material for top cells in tandem photovoltaic configurations. Nevertheless, the prominence of BaZrS$_3$ also underscores the limited experimental diversity currently available within this materials family. 

Although theoretical studies predict that many chalcogenide perovskites possess favorable optoelectronic properties \cite{basera2022_chalcogenide_perovskites,chakravorty2025_unlocking_opt, adhikari2025_leadfree_cp, adhikari2025_sn_cp}, experimental realization of the perovskite-type phase remains scarce beyond a few compositions, including BaHfS$_3$~\cite{nishigaki2020}, Sr(Hf,Zr)S$_3$~\cite{liang2023}, and EuZrS$_3$~\cite{lelieveld1980}. Moreover, the synthesis of these materials typically requires high-temperature processes, raising concerns for device integration and scalability. As a result, significant experimental effort has been directed toward lowering synthesis temperatures and identifying viable processing routes~\cite{comparotto2022, comparotto2025}. Many of these challenges arise from the high thermodynamic stability of binary and polysulfide phases, as well as the presence of competing non-perovskite structures, which can hinder the formation of the perovskite-type phase. This disconnect between theoretical predictions and experimental feasibility represents a major bottleneck in the discovery of new chalcogenide perovskites.

The perovskite-type structure consists of a corner-sharing $BX_6$ octahedral network with $A$ cations occupying the dodecahedral sites \cite{Breternitz2018, Rose1840}. To rationalize perovskite phase stability, simple geometrical descriptors such as the Goldschmidt tolerance factor ($t_{\text{Goldschmidt}}$) and the octahedral factor ($\mu$) have long been employed based on ionic radii ($r_A$, $r_B$, and $r_X$) \cite{goldschmidt1926_krystallochemie}. These dimensionless parameters predict perovskite formation when their values fall within the approximate ranges of 0.8-1 and 0.414-0.732, respectively:

\begin{equation}
    t_{\text{Goldschmidt}} = \frac{r_A + r_X}{\sqrt{2}(r_B + r_X)} \quad
\quad \mu = \frac{r_B}{r_X}
\end{equation}

However, these classical descriptors fail to reliably predict phase stability in chalcogenide perovskites, as several non-perovskite chalcogenide phases also fall within these nominal stability ranges~\cite{Jess2022, turnley2024}. This limitation has been attributed to the increased covalent character of metal-chalcogen bonds compared to oxide or halide analogues. To address this issue, modified tolerance factors incorporating electronegativity differences have been proposed, including the Jess \textit{et al.} tolerance factor ($t_{\text{Jess}}$) and  electronegativity-based descriptors like ${\chi_\mathrm{diff}}$~\cite{Jess2022, turnley2024}.

\begin{equation}
    t_{\text{Jess}} = \frac{\frac{\Delta\chi_{A-X}}{\Delta\chi_{A-O}} (r_A + r_X)}{\sqrt{2} \frac{\Delta\chi_{B-X}}{\Delta\chi_{B-O}} (r_B + r_X)} 
\end{equation}

While these descriptors improve discrimination between perovskite and non-perovskite chalcogenide phases, their generality and predictive power across unexplored compositional spaces remain uncertain, motivating the development of more systematic screening approaches.

Alternative screening methodologies rely on high-throughput first-principles calculations, where crystal structures are relaxed and materials properties such as thermodynamic stability, electronic band structure, and optical response are computed~\cite{huo2018_highthroughput, cao2024_highthroughput, singh2025_amse3}. The rise of these computational approaches has significantly impacted the identification of lead-free perovskites, particularly in data-scarce regimes where experimental exploration is costly and time-consuming. When coupled with machine-learning techniques, these methods enable rapid exploration of large chemical spaces. However, experimental validation of computational predictions remains limited, and structures predicted to be stable by density functional theory (DFT) do not always correspond to experimentally realizable phases~\cite{fatima2025_leadfreeperovskites}.

In parallel with high-throughput first-principles approaches, materials informatics has evolved rapidly over the past decade, enabling property prediction directly from composition and/or crystal structure. Early efforts focused on descriptor engineering and sparse regression frameworks such as SISSO (sure independence screening and sparsifying operator), which identify compact, physically interpretable analytical expressions from large feature spaces \cite{ouyang2018_sisso}. More recently, deep learning models such as CrabNet and MODNet have demonstrated accurate composition-based property prediction without explicit structural input \cite{Wang2021crabnet, de_breuck_materials_2021}. In addition, generative models for crystal structure prediction have emerged \cite{de_breuck_generative_2025},  ranging from variational autoencoders and diffusion-based approaches to large language models capable of generating complete crystallographic information files \cite{antunes_crystal_2024}. These advances increasingly enable not only forward property prediction but also constraint-aware candidate generation and screening, expanding the scope of data-driven materials discovery beyond conventional paradigms.

\begin{figure*}[ht]
  \centering
  \includegraphics[width=\textwidth]{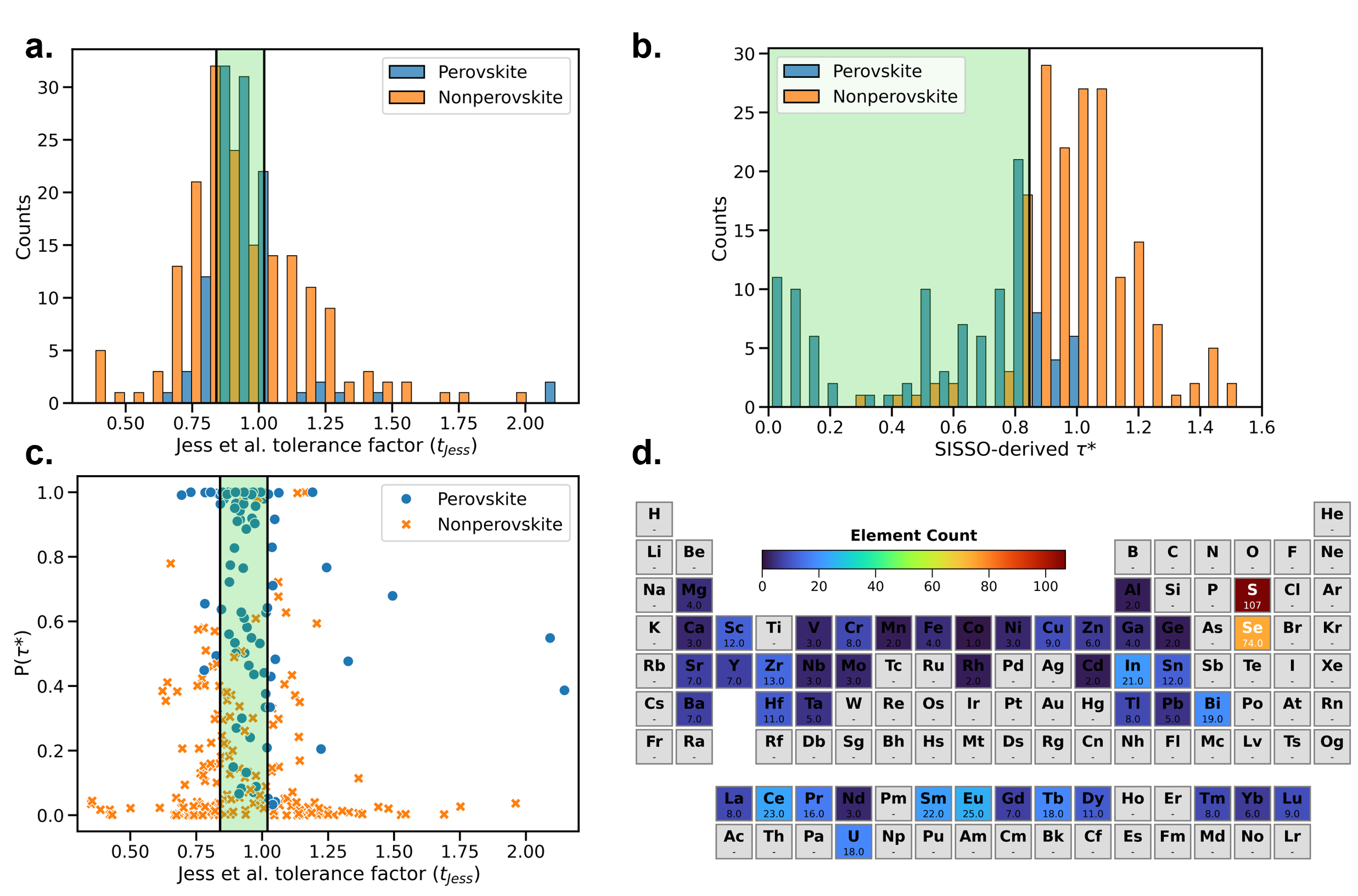}
\caption{
a. Jess \textit{et al.} tolerance factor ($t_{\text{Jess}}$) distribution on the experimental ABX$_3$ data used for training;
b. SISSO-derived $\tau^*$ tolerance factor distribution on the experimental ABX$_3$ data;
c. Logistic-calibrated probability of perovskite-type stability based on $\tau^*$ as a function of the Jess \textit{et al.} tolerance factor for the experimental ABX$_3$ data;
in each plot the stability region is delimited with a green background;
d. Elemental distribution of the predicted perovskite-type phases with $\tau^* < 0.846$.
}
  \label{fig:data_distribution}
\end{figure*}

Here, we present a machine-learning-guided screening pipeline aimed at identifying potentially synthesizable chalcogenide perovskites, with an emphasis on experimentally grounded descriptors (Figure~\ref{fig:workflow}). A new analytical tolerance factor derived using the SISSO framework is introduced to assess structural stability. Sequentially, crystal structure, experimental plausibility, and bandgap are inferred using machine-learning models, including CrystaLLM, CrabNet, GCNNs~\cite{antunes_crystal_2024,gu2022_perovskite_synth, Wang2021crabnet}. The resulting ranked list of candidate materials is intended to provide a focused platform for experimental validation and to stimulate discussion on the strengths, limitations, and future directions of ML-driven materials identification in chalcogenide perovskites. Notably, each stage of the pipeline introduces its own uncertainty; the ensemble therefore does not remove uncertainty but distributes decision-making across partially independent screening criteria.

\section{Results and Discussion}

\subsection{ML-derived tolerance factor}

The tolerance factor $\tau^*$ was derived using the SISSO algorithm based on experimentally reported halide and chalcogenide perovskites~\cite{ouyang2018_sisso}. The resulting expression for $\tau^*$ represents an analytically compact descriptor that balances model sparsity with predictive accuracy, while retaining direct physical interpretability in terms of geometric packing and size mismatch. $\tau^*$ (Eq. 3) is a function of the ionic radii, where $r_A$, $r_B$, and $r_X$ denote the ionic radii of the A-site cation, B-site cation, and anion, respectively. Perovskite stability is defined by the criterion $\tau^* < 0.846$, which was learned from the experimental data with a decision tree classifier (see \nameref{sec:data}).

\begin{equation}
\tau^* = \left| \frac{r_A + r_B}{r_X}
+ \left| \frac{r_B}{r_X} - \ln\!\left( \frac{r_A}{r_B} \right) \right|
- \left( \frac{r_A}{r_X} \right)^3 \right| ,
\end{equation}

Following Turnley \textit{et al.}, we used their chalcogenide-tailored ionic radii for larger and more polarizable anions (S$^{2-}$, Se$^{2-}$, Br$^-$, I$^-$), while Shannon ionic radii were retained for the more ionic F$^-$ and Cl$^-$ systems \cite{turnley2024}. This choice reflects the increased covalent character of metal–chalcogen bonds compared to oxide and fluoride analogues. We note that the selection of ionic radii constitutes an important modeling assumption, and further refinement of radii datasets for specific ABX$_3$ chemistries may influence the quantitative form of the derived descriptor.

\begin{table}[H]
    \centering
    \caption{Classification performance metrics for the Goldschmidt ($t_{\text{Goldschmidt}}$), Jess \textit{et al.} ($t_{\text{Jess}}$), and Bartel \textit{et al.} ($t_{\text{Bartel}}$) tolerance factors together with the SISSO-derived $\tau^*$ for perovskite stability prediction.}
    \label{tab:model_performance}
    \begin{tabular}{lcccc}
    \hline
    Metric & $t_{\text{Goldschmidt}}$ & $t_{\text{Jess}}$ & $t_{\text{Bartel}}$ & $\tau^*$ \\
    \hline
    Accuracy (\%)  & 59.6 & 70.2 & 70.2 & \textbf{91.2} \\
    Precision (\%) & 50.0 & 59.4 & 59.4 & \textbf{95.0} \\
    Recall (\%)    & 60.9 & \textbf{82.6} & \textbf{82.6} & \textbf{82.6} \\
    Specificity (\%) & 58.8 & 61.8 & 61.8 & \textbf{97.1} \\
    F1-score (\%)  & 55.1 & 69.1 & 69.1 & \textbf{88.4} \\
    \hline
    \end{tabular}
\end{table}

\begin{figure*}[ht]
  \centering
  \includegraphics[width=\textwidth]{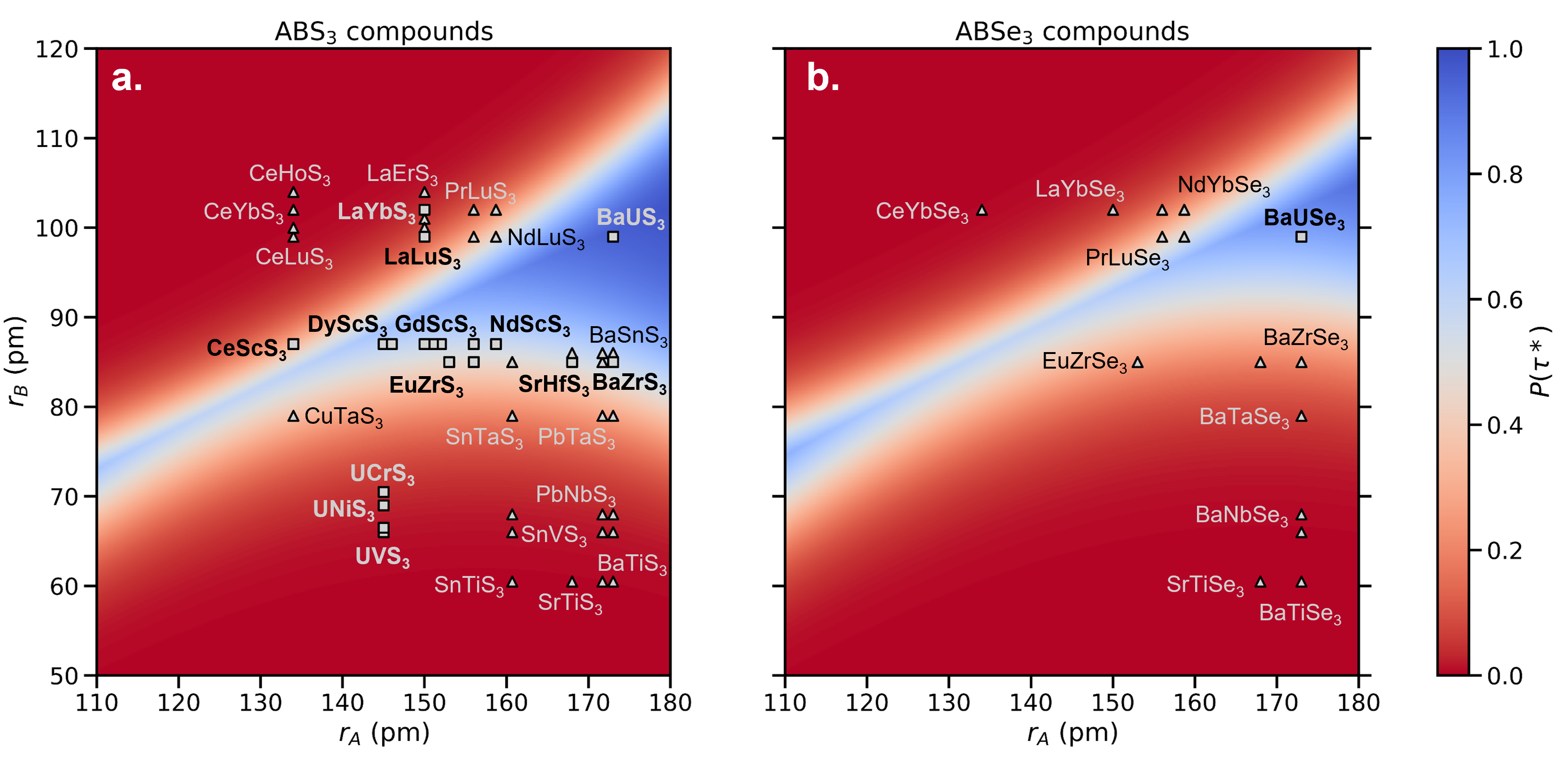}
\caption{The effects of ionic radii on the stability of chalcogenide perovskites for a. ABS$_3$ and b. ABSe$_3$ compounds. The scattered symbols correspond to experimentally synthesized materials, materials with a perovskite-type structure are represented with squares and a bold label, while non-perovskite structures are triangles.}
  \label{fig:colormap_radii}
\end{figure*}

Figure~\ref{fig:data_distribution} shows the distribution of the experimental dataset for the reported Jess \textit{et al.} tolerance factor ($t_{\text{Jess}}$) \cite{Jess2022} (a) and the SISSO-derived $\tau^*$ introduced in this work (b). In addition, inspired by Bartel \textit{et al.}, we fitted a logistic calibration that maps $\tau^*$ values to a probability of perovskite-type stability derived from the experimental phase labels \cite{Bartel19}. Figure~\ref{fig:data_distribution}c shows the calibrated probability as a function of $t_{\text{Jess}}$. Notably, most experimentally observed perovskites—both within and outside the nominal $t_{\text{Jess}}$ stability window—exhibit calibrated probabilities above 50\%.

Table \ref{tab:model_performance} compares the derived $\tau^*$ with already established descriptors like $t_{\text{Jess}}$ \cite{Jess2022}, Goldschmidt tolerance factor ($t_{\text{Goldschmidt}}$) and Bartel tolerance factor ($t_{\text{Bartel}}$) \cite{Bartel19}, while Fig.~S\ref{fig:cm_tolerance_factor} shows the respective confusion matrices. The four formulae exhibit distinct predictive performance in terms of accuracy and F1-score for the experimental test dataset. $t_{\text{Jess}}$ and $t_{\text{Bartel}}$ achieve identical results in terms of accuracy and F1-score, while the baseline model $t_{\text{Goldschmidt}}$ performs worst, reaching an F1-score of 55.1\%. In contrast, $\tau^*$ clearly outperforms all other approaches, attaining the highest accuracy and F1-score.

$t_{\text{Jess}}$, $t_{\text{Bartel}}$, and $\tau^*$ exhibit comparable recall values (82.6\%), indicating a similar ability to correctly identify experimentally stable perovskites. However, $t_{\text{Jess}}$ and $t_{\text{Bartel}}$ produce a substantial number of false positives, corresponding to unstable perovskite-type compositions incorrectly classified as stable; such errors are particularly undesirable, as they can lead to unnecessary synthesis and characterization efforts.

$\tau^*$ significantly mitigates this limitation by reducing the number of false positives to a single case, while preserving the same recall as $t_{\text{Jess}}$ and $t_{\text{Bartel}}$. This improvement results in a marked increase in precision (95.0\%) and specificity (97.1\%), indicating a much higher reliability of stability predictions. Figure~\ref{fig:data_distribution}d shows the elemental distribution of the 181 predicted perovskites from the full chemical space of 1392 possible ABX$_3$ chalcogenide compositions. This low fraction is consistent with experimental observations, as only a limited number of chalcogenide compounds have been reported to crystallize in the perovskite-type structure, with many candidate compositions instead favoring non-perovskite phases. As expected, more sulfur compounds are predicted to be stable than selenium-based materials, consistent with the larger ionic radius of Se. Predictions for ABSe$_3$ compositions should therefore be interpreted with caution, as experimentally reported selenide perovskites are underrepresented in the training dataset, limiting the statistical confidence of the learned stability boundary for this subset. In addition, common observed elements in the chalcogenide perovskite literature are well represented including Hf, Zr, Ba, Eu, and U.

\begin{figure*}[ht]
  \centering
  \includegraphics[width=\textwidth]{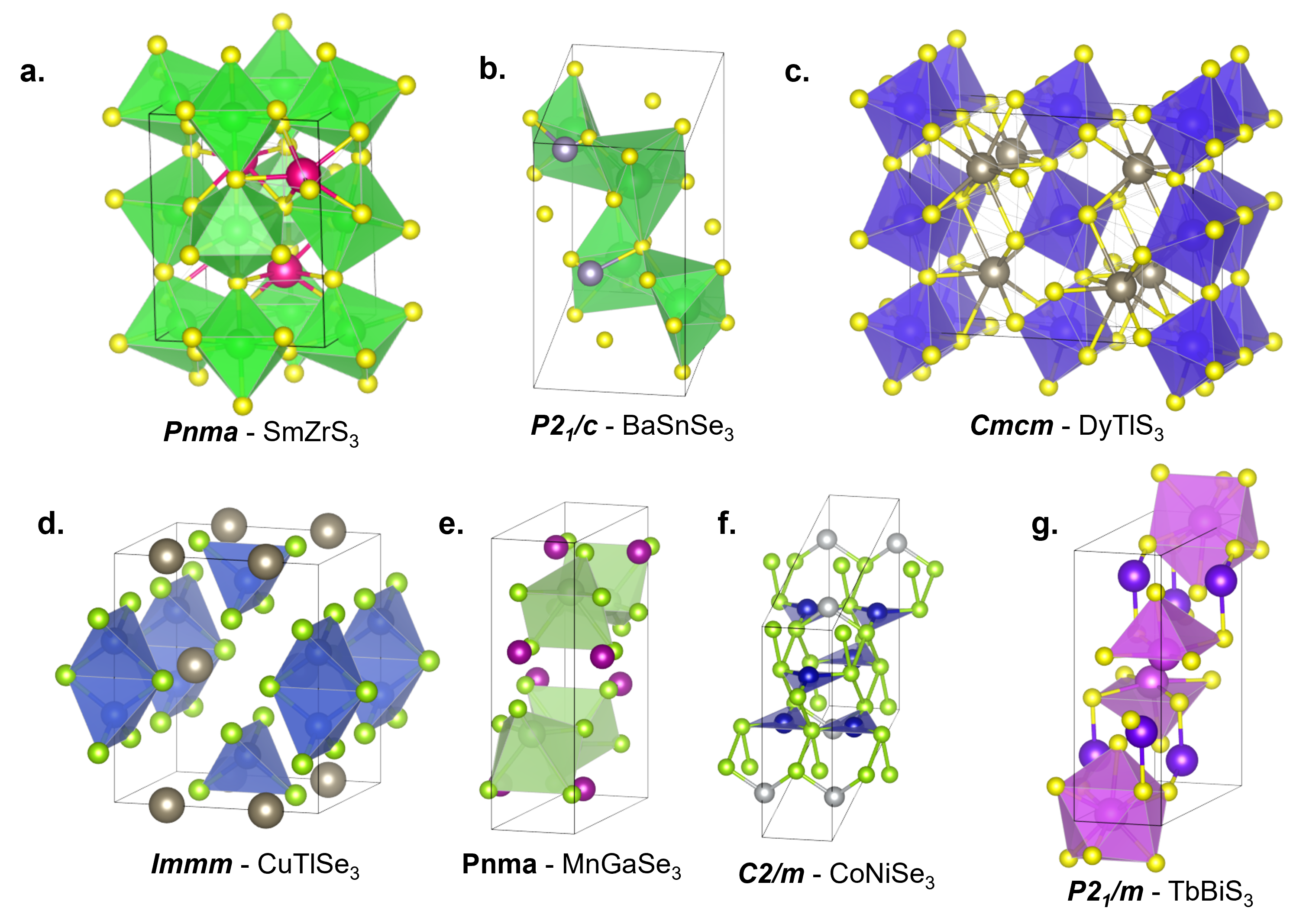}
\caption{Some of the generated crystal structures with \textit{CrystaLLM}, and their respective space group, based on the formulae that are predicted stable with $\tau^*$ including some specific compounds as examples: a. SmZrS$_3$ , b. BaSnSe$_3$ c. DyTlS$_3$, d. CuTlSe$_3$, e. MnGaSe$_3$, f. CoNiSe$_3$, g. TbBiS$_3$. Yellow circles correspond to the chalcogenide anions, while the other colors are for the A and B cations.}
  \label{fig:crys_structures}
\end{figure*}

Unlike the classical Goldschmidt tolerance factor, $\tau^*$ incorporates higher-order and mismatch-dependent terms. A defining feature of  $\tau^*$ is its intrinsic asymmetry with respect to the A- and B-site cations. The A-site radius $r_A$ enters both linearly and through a dominant cubic term $(r_A/r_X)^3$, while the B-site radius $r_B$ enters linearly and logarithmically via a size-mismatch term. This asymmetry reflects the distinct structural roles of the two cation sites: the A-site cation controls the size and rigidity of the anion framework, whereas the B-site cation primarily determines the dimensions of the corner-sharing BX$_6$ octahedra.

When evaluated over the entire chemically accessible range of ionic radii, the $\tau^*$ expression admits two analytic branches, separated by the condition $\frac{r_B}{r_X} = \ln(\frac{r_A}{r_B})$. On one branch, $\tau^*$ retains an explicit linear dependence on $r_B$; on the other, the linear B-site contributions cancel, leaving only a logarithmic dependence on the B-site radius. Figure~\ref{fig:colormap_radii} shows color maps of $\tau^*$ across the full $(r_A,r_B)$ chemical space for both anions. In our screened radius range, the stability region occurs primarily for $r_A > r_B$, which lies within the latter branch. Consequently, although linear B-site contributions are present in the general mathematical form of $\tau^*$, they largely cancel within the stability-allowed region. 

Within the stability-allowed region of the training and screening data, $\tau^*$ exhibits stronger sensitivity to the A-site ionic radius, the cubic term $(r_A/r_X)^3$ introduces a sharp upper bound on $r_A$ in the descriptor, beyond which $\tau^*$ increases rapidly. In contrast, variations in the B-site radius contribute more gradually through the logarithmic mismatch term. As a result, $r_B$ acts as a fine-tuning parameter, while $r_A$ largely governs whether a perovskite-type structure is geometrically feasible within this model. The resulting stability region defined by $\tau^*$ therefore takes the form of a narrow wedge in the $(r_A,r_B)$ space. Fig.~S\ref{fig:matrix_prob} shows an elemental matrix color-coded with the logistic-calibrated stability probability derived from $\tau^*$.

\subsection{Structural validation}

To further improve the precision of our screening, we employ an ensemble strategy that combines the $\tau^*$ criterion with explicit structure generation using \textit{CrystaLLM}. While $\tau^*$ identifies compositions that are geometrically and chemically plausible perovskites, \textit{CrystaLLM} provides an independent assessment of whether these compositions can realistically adopt a perovskite-type structure based solely on their chemical formula, without resorting to computationally expensive first-principles calculations \cite{antunes_crystal_2024}.

Crystal structures were generated for each of the 181 chalcogenide compositions satisfying $\tau^* < 0.846$. At this stage, density functional theory (DFT) verification was intentionally not performed, as \textit{CrystaLLM} is trained on DFT-derived structures \cite{antunes_crystal_2024}. Consequently, the generated structures are not guaranteed to correspond to fully relaxed energetic minima at the DFT level of theory. These generated structures (see \nameref{sec:data}) were further analyzed to determine whether they adopted the true perovskite-type topology characterized by a corner-sharing $BX_6$ octahedral network. Non-perovskite structures were filtered by analyzing the a–b–c lattice parameters of the generated crystal structures, where clusters corresponding to edge-sharing and corner-sharing motifs within the \textit{Pnma} space group were observed.

\begin{figure*}[ht]
  \centering
  \includegraphics[width=\textwidth]{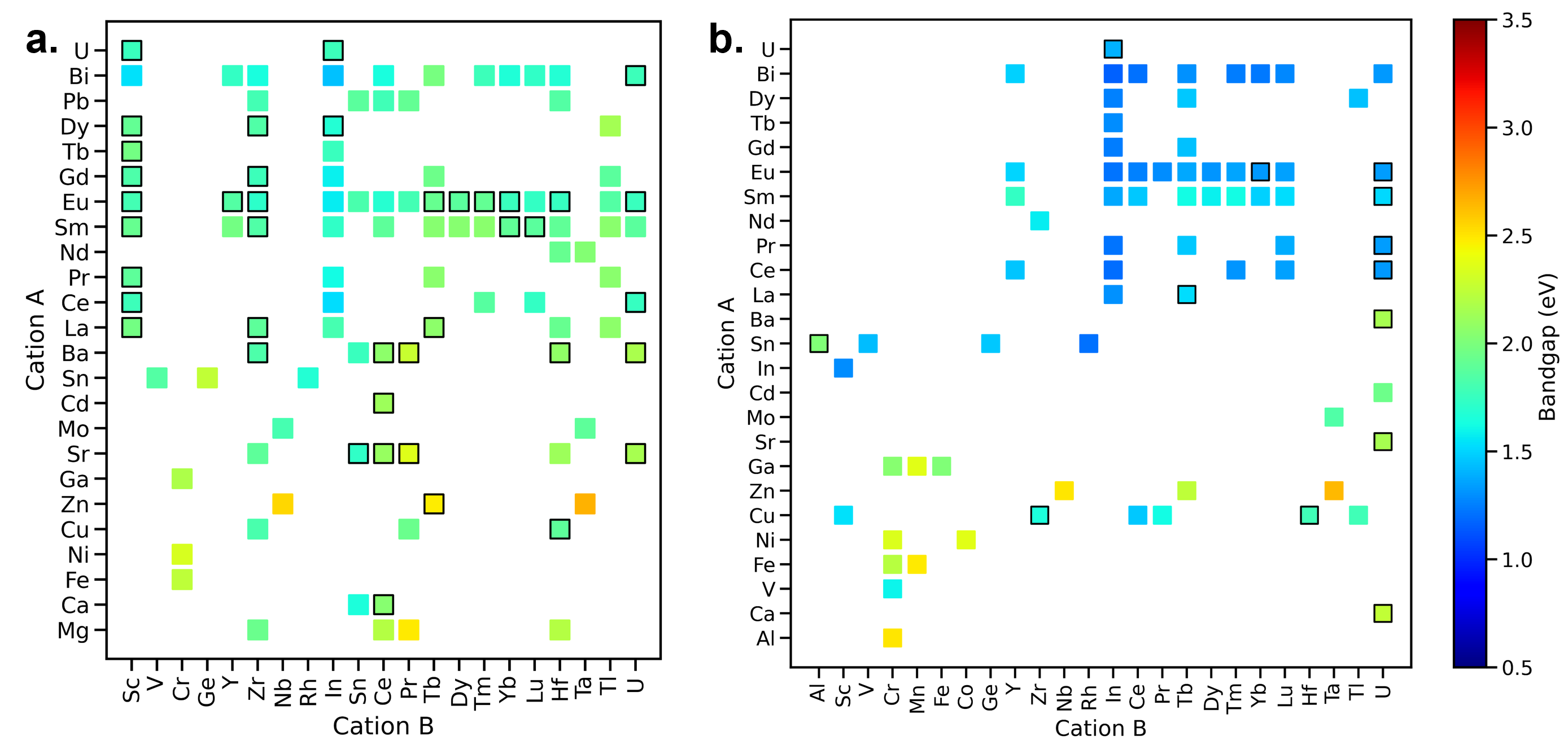}
  \caption{CrabNet-estimated bandgaps for predicted chalcogenide perovskites displayed on an element–element matrix for a. ABS$_3$ and b. ABSe$_3$ compositions. Color indicates the predicted bandgap value, while squares outlined in black correspond to compounds predicted to adopt a corner-sharing perovskite-type structure according to \textit{CrystaLLM}.}
  \label{fig:matrix_eg}
\end{figure*}

Requiring agreement between both $\tau^*$ and \textit{CrystaLLM} models  reduces the candidate set to 54 compositions, corresponding to approximately 30\% of the $\tau^*$-stable subset and about 4\% of the original chemical space of 1392 possible compounds. This reduction reflects an intentional trade-off between recall and precision, because our ensemble prioritizes candidates supported by both geometric and structural hypotheses, giving a more experimentally actionable set of materials. Higher precision, or confidence in the predicted candidates, is favorable for screening tasks, especially for this type of materials as synthesis efforts have reported challenges related to byproducts and formation of non-perovskite phases~\cite{carr2025_origins, Sopiha_2022}. In addition, Fig.~S\ref{fig:crystal_ICSD} presents parity comparisons of the lattice parameters ($a$, $b$, $c$) and unit-cell volume between the generated structures and ICSD-reported experimental data, showing close quantitative agreement.

The distributions of normalized ionic radii for the 54 \textit{CrystaLLM}-validated perovskites reveal a significantly narrower stability window than the one defined by the tolerance factor alone. The A-site ratio $r_A/r_X$ is strongly concentrated toward large values, with a mean of $0.84 \pm 0.09$ ($r_A = 155\pm17$~pm for ABS$_3$ compounds), indicating that successful formation of chalcogenide perovskites requires near-optimal filling of the anion framework by the A-site cation. Only a small number of compounds exhibit substantially smaller A-site cations, and the upper bound of the distribution approaches the geometric cutoff imposed by the cubic A-site term in $\tau^*$. In contrast, the B-site ratio $r_B/r_X$ is confined to a narrow interval centered around $0.48 \pm 0.04$ ($r_B = 88\pm7$~pm for ABS$_3$ compounds). Both smaller and larger B-site cations are systematically eliminated during structure generation, despite being permitted by the tolerance-factor criterion, indicating that only a limited range of octahedral sizes is compatible with stable corner-sharing connectivity in chalcogenide systems.

Further insight into this refinement is obtained by examining the space-group symmetries of the 181 $\tau^*$-stable compounds. Figure~\ref{fig:crys_structures} shows some of the generated structures, which span several low-symmetry space groups, including 116 in \textit{Pnma}, 46 in \textit{Cmcm}, 11 in \textit{C2/m}, 4 in \textit{P2$_1$/c}, 3 in \textit{P2$_1$/m}, and 1 in \textit{Immm}. The orthorhombic \textit{Pnma} space group dominates this set, consistent with its well-established role as a common distorted derivative of the ideal cubic perovskite-type structure, either as corner- or edge-sharing octahedral networks.

These results demonstrate that the $\tau^*$ tolerance factor efficiently narrows the chemical space to geometrically plausible candidates, while \textit{CrystaLLM} structure generation resolves topological competition and isolates the small subset of compounds that reliably realize the perovskite-type structure. These 54 candidates include already reported compounds like AScS$_3$ (A = Ce, Dy, Gd, La, Pr, Sm, Tb), BaBS$_3$ (B = Hf, U, Zr), EuBS$_3$ (B = Hf, Zr) and SrSnS$_3$, which is reported as a non-perovskite phase \cite{yamaoka1970_baSnS3}. Interestingly, there are computational studies that describe the potential of this material in the perovskite-type structure as an absorber layer in solar cells \cite{adhikari2025_sn_cp, ju2017_perovskite_chalcogenides}. Moreover, new chalcogenide perovskites are predicted to form stable perovskite-type phases including several Uranium-based compounds, AZrS$_3$ (A = Dy, Gd, La, Sm), EuBS$_3$ (B$^{3+}$ = Dy, Tb, Tm, Yb), BaCeS$_3$, EuScS$_3$, DyInS$_3$, among others.

\subsection{Suitability for PV applications}

\begin{figure*}[ht]
  \centering
  \includegraphics[width=\textwidth]{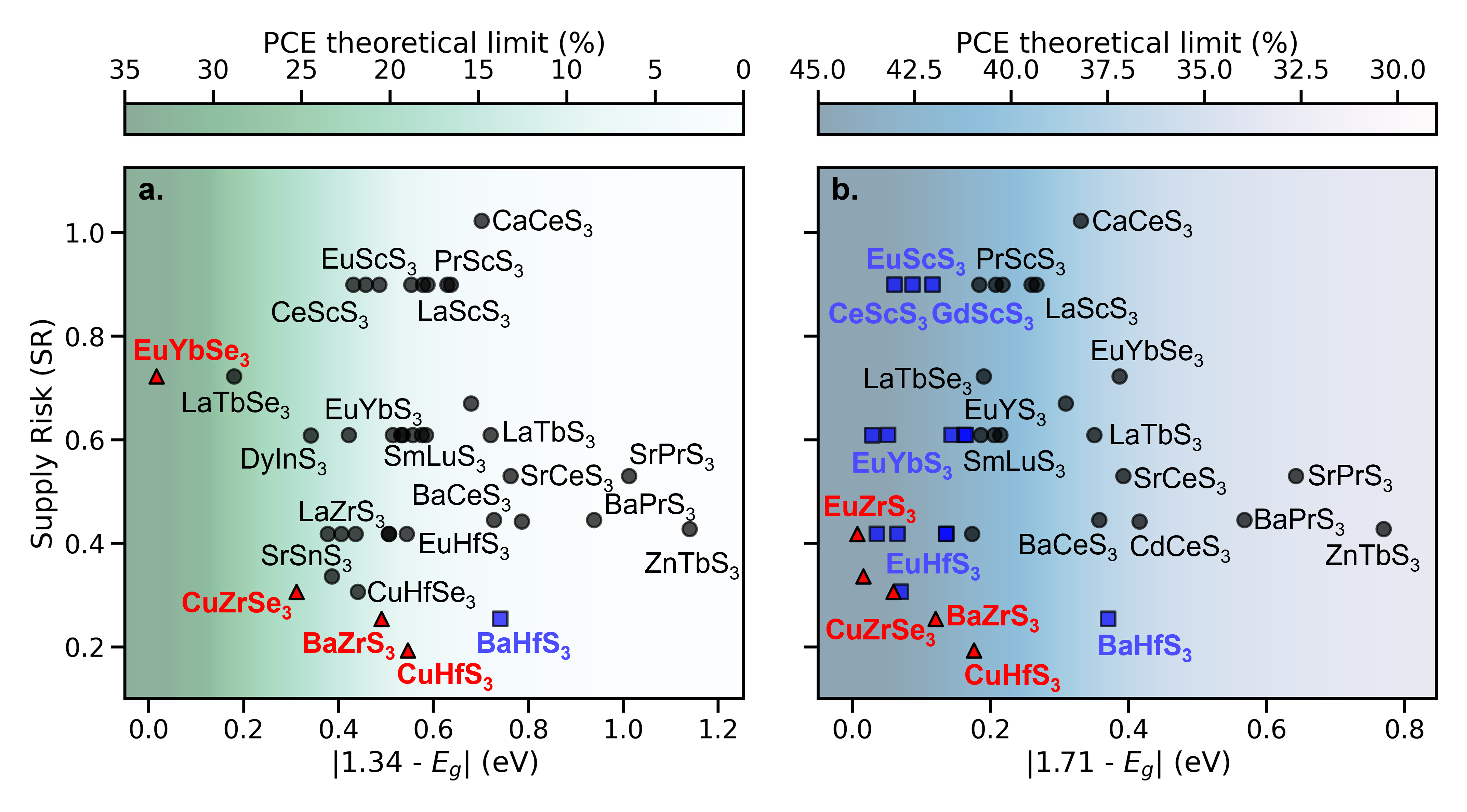}
  \caption{
  Evaluation of material sustainability using the supply risk (SR), derived from the Herfindahl-Hirschman Index (HHI) and ESG scores, as a function of the deviation of the estimated bandgap from the optimal values for (a) single-junction photovoltaics ($E_g^{opt} = 1.34$~eV) and (b) the top cell in a tandem photovoltaic configuration ($E_g^{opt} = 1.71$~eV).
  Red triangles indicate Pareto-optimal materials that simultaneously minimize the bandgap deviation $|E_g - E_g^{opt}|$ and SR.
  Blue squares correspond to materials lying within 10\% of the minimum values of both optimization objectives. The background color indicates the theoretical limit in power conversion efficiency (PCE) based on the Shockley-Queisser limit~\cite{shockley1961_detailedbalance}.}
  \label{fig:sr_bandgap}
\end{figure*}

To assess the suitability of the 54 candidates for PV applications, two indicators were studied. Firstly, the bandgap of each material was estimated using a composition-based regression algorithm \cite{Wang2021crabnet}. Secondly, a sustainability factor was estimated with indices based on the elemental world production (Herfindahl-Hirschman Index, HHI) and environmental, social and governance (ESG) scores. Inspired by Nominé \textit{et al.} a supply risk was determined for each composition \cite{nomine2023}.

Bandgap estimation was possible by training a CrabNet model based on reported experimental data from halide perovskites, chalcogenide semiconductors, and chalcogenide perovskites. This dataset was chosen to have an equilibrium between the perovskite-type structure knowledge and the chalcogenide effect. In addition, training solely with experimental chalcogenide perovskites was not feasible as the reported data is limited. The influence of training dataset size on model performance is shown in Fig.~S\ref{fig:dataset_size}, illustrating convergence of the mean absolute error as the dataset grows.

Evaluation on the held-out test set yielded a mean absolute error (MAE) of 248~meV for the bandgap. The corresponding parity plot is shown in Fig.~S\ref{fig:bg_parity}. This is consistent with the validation performance and demonstrates stable generalization to unseen compositions drawn from the same chemical distribution. Notably, the test MAE did not show a significant change when using different element encoders, including Magpie~\cite{ward2016_magpie}, Mat2Vec~\cite{tshitoyan2019_mat2vec}, one-hot, Pettifor~\cite{Cerqueira2025} descriptors and even randomly initialized embeddings (Fig.~S\ref{fig:comp_encoders}). The near-invariance of the final performance to the choice of encoder indicates that CrabNet’s attention mechanism learns chemically meaningful element representations directly from composition during training, rapidly overriding the specific priors encoded in the initial embeddings \cite{wang2021_crabnet_explainable}. In this regime, the dominant limitation is not feature expressiveness but rather the absence of structural information and the inherent noise in experimental bandgap measurements, which depend on synthesis conditions, measurement techniques, temperature, defects, and sample morphology.

Importantly, the achieved MAE falls within the range typically reported for existing benchmarks for composition-only prediction of experimental bandgaps. Prior studies on standardized benchmarks typically report MAEs in the range of 0.3–0.5 eV for the \texttt{Matbench v0.1} experimental dataset when crystal structure information is not considered  \cite{dunn2020_matbench, debreuck2022_bandgap, vasylenko2025_digital_features}. We emphasize that the absolute MAE obtained for property estimation is extremely dataset dependent, reflecting differences in chemical space, data curation, and experimental uncertainty. As such, the comparison to prior benchmarks is not intended as a strict performance evaluation, but rather to contextualize the present results within the typical accuracy range reported for composition-only models trained on experimental bandgap data. 

Figure~\ref{fig:matrix_eg} shows elemental matrices with the predicted bandgap color coded for all $\tau^*$ stable compounds, with a black edge for the \textit{CrystaLLM} predicted perovskite-type structures. Narrow bandgaps (blue squares in Fig.~\ref{fig:matrix_eg}) are largely filtered out by the \textit{CrystaLLM} step. As a result, the subset of compounds classified as perovskite-type structures exhibits a bandgap distribution with a mean of 1.86~eV. This trend is consistent with literature reports, as many chalcogenide perovskites exhibit relatively wide bandgaps. For example, the well-known chalcogenide perovskites BaZrS$_3$ and BaHfS$_3$ are predicted to have bandgaps of $1.8 \pm 0.4$ eV and $2.1 \pm 0.5$ eV, respectively, in agreement with reported experimental values \cite{Sopiha_2022, agarwal2025_bazrs3_review, vincent2023_liquidflux, nishigaki2020}. In contrast, the predicted bandgap of LaScS$_3$ ($2.0 \pm 0.5$ eV) underestimates a recently reported experimental value of 2.9 eV \cite{zhang2025_ptccm_adfm}, even though it was included in the training dataset. While a single discrepancy does not allow for a quantitative assessment of model accuracy, these results support the use of the model as a rapid screening tool for identifying photovoltaic-relevant candidates rather than as a substitute for high-level theoretical or experimental methods. Notably, this experimental study does not report an error associated with the measurement, and strong absorption is observed in the data at photon energies approaching 2 eV \cite{zhang2025_ptccm_adfm}. Thus further corroboration of this value may be required.

To further illustrate the chemical origin of the CrabNet bandgap predictions, we performed a principal component analysis (PCA) shown in Fig.~S\ref{fig:pca_analysis}. The first two principal components, which together account for more than 60\% of the bandgap variance, are primarily governed by descriptors associated with the B-site cation and the chalcogenide anion, emphasizing the dominant role of the BX$_6$ octahedral framework in defining the electronic structure of chalcogenide perovskites. This trend is consistent with established electronic-structure models of perovskites, in which the valence and conduction band edges are primarily derived from X-anion p states and B-cation d states states, respectively, while the A-site cation mainly influences the bandgap indirectly through structural distortions and lattice effects \cite{Sopiha_2022}.

\subsection{Sustainability analysis}

Figure~\ref{fig:sr_bandgap} illustrates the combined influence of optoelectronic performance and sustainability by plotting the supply risk (SR) against the deviation of the predicted bandgap from optimal photovoltaic values. Two device architectures were considered. For single-junction photovoltaics, the optimal bandgap was taken as $E_g^{opt} \approx 1.34$~eV, corresponding to the maximum theoretical efficiency predicted by the Shockley-Queisser limit~\cite{shockley1961_detailedbalance}. In contrast, the typically wider bandgaps of chalcogenide perovskites make them more suitable as absorbers for top cells in tandem photovoltaic architectures. In this configuration, the top cell absorbs high-energy photons while remaining transparent to lower-energy photons transmitted to the bottom cell, thereby surpassing the single-junction efficiency limit. For tandem devices paired with silicon bottom cells, the optimal bandgap of the top absorber was taken as $E_g^{opt} = 1.71$~eV~\cite{fell2025_tandem_efficiency}. Uranium-based compounds were excluded from this analysis due to their inherent toxicity and radioactivity, which prevent practical photovoltaic applications.

Pareto-optimal candidates (red triangles) identify materials that achieve the best compromise between electronic suitability and sustainability. Notably, the well-studied compound BaZrS$_3$ remains optimal for tandem photovoltaic configurations, consistent with its frequent identification as a benchmark absorber in prior experimental studies. In addition, several previously unexplored compositions, including CuHfS$_3$, CuZrSe$_3$, and EuYbSe$_3$, emerge as promising photovoltaic absorbers. Materials highlighted in blue (squares) delineate a practically attractive region of the design space, lying within 10\% of the minimum values for both optimization objectives. This subset exhibits increased chemical diversity, including compounds containing rare-earth elements, consistent with their large ionic radii and favorable geometric compatibility with the perovskite-type structure. While materials such as EuYbS$_3$, CeScS$_3$, and EuScS$_3$ may be of fundamental interest, their relatively high supply risk may limit their feasibility for large-scale photovoltaic deployment.

\subsection{Assessment of experimental plausibility}

\begin{table*}[t!]
\centering
\caption{Data derived for \textit{CrystaLLM}-predicted stable, true perovskite-type compositions, excluding U-containing compounds. The compounds are sorted by CLS and the bold compounds have already been seen experimentally. HHI corresponds to the Herfindahl-Hirschman Index, SR is the Supply Risk. The bandgap and its standard deviation were predicted using a trained CrabNet model, while CLS is the crystal-likeness score, which is a data-driven proxy for relative synthesizability derived from a pre-trained GCNN model.}
\label{tab:all_comp}
\begin{tabular}{l c c c c c}
\toprule
Formula & $\tau^*$ & Bandgap (eV) & HHI & SR & CL score \\
\midrule
\textbf{BaZrS$_3$}   & \textbf{0.820} & \textbf{$1.830 \pm 0.463$} & \textbf{0.485} & \textbf{0.254} & \textbf{$0.991 \pm 0.024$} \\
EuScS$_3$   & 0.817 & $1.797 \pm 0.491$ & 1.586 & 0.899 & $0.988 \pm 0.019$ \\
\textbf{BaHfS$_3$}   & \textbf{0.820} & \textbf{$2.081 \pm 0.511$} & \textbf{0.485} & \textbf{0.254} & \textbf{$0.987 \pm 0.025$} \\
\textbf{DyScS$_3$}   & \textbf{0.809} & \textbf{$1.917 \pm 0.377$} & \textbf{1.586} & \textbf{0.899} & \textbf{$0.985 \pm 0.017$} \\
\textbf{CeScS$_3$}   & \textbf{0.821} & \textbf{$1.771 \pm 0.442$} & \textbf{1.586} & \textbf{0.899} & \textbf{$0.982 \pm 0.031$} \\
\textbf{EuZrS$_3$}   & \textbf{0.840} & \textbf{$1.717 \pm 0.417$} & \textbf{0.778} & \textbf{0.418} & \textbf{$0.981 \pm 0.030$} \\
SrSnS$_3$   & 0.822 & $1.726 \pm 0.385$ & 0.581 & 0.336 & $0.981 \pm 0.036$ \\
\textbf{GdScS$_3$}   & \textbf{0.820} & \textbf{$1.826 \pm 0.347$} & \textbf{1.586} & \textbf{0.899} & \textbf{$0.981 \pm 0.043$} \\
\textbf{LaScS$_3$}   & \textbf{0.818} & \textbf{$1.969 \pm 0.505$} & \textbf{1.586} & \textbf{0.899} & \textbf{$0.975 \pm 0.082$} \\
\textbf{EuHfS$_3$}   & \textbf{0.840} & \textbf{$1.745 \pm 0.429$} & \textbf{0.778} & \textbf{0.418} & \textbf{$0.973 \pm 0.067$} \\
GdZrS$_3$   & 0.844 & $1.775 \pm 0.325$ & 0.778 & 0.418 & $0.971 \pm 0.091$ \\
EuTmS$_3$   & 0.775 & $1.916 \pm 0.343$ & 1.076 & 0.609 & $0.959 \pm 0.111$ \\
EuTbS$_3$   & 0.757 & $1.924 \pm 0.302$ & 1.076 & 0.609 & $0.957 \pm 0.106$ \\
\textbf{TbScS$_3$}   & \textbf{0.819} & \textbf{$1.976 \pm 0.349$} & \textbf{1.586} & \textbf{0.899} & \textbf{$0.956 \pm 0.068$} \\
LaZrS$_3$   & 0.841 & $1.883 \pm 0.462$ & 0.778 & 0.418 & $0.956 \pm 0.114$ \\
LaTbS$_3$   & 0.793 & $2.060 \pm 0.225$ & 1.076 & 0.609 & $0.955 \pm 0.100$ \\
DyZrS$_3$   & 0.833 & $1.845 \pm 0.350$ & 0.778 & 0.418 & $0.950 \pm 0.125$ \\
EuYS$_3$    & 0.796 & $1.854 \pm 0.301$ & 1.076 & 0.609 & $0.944 \pm 0.109$ \\
LaTbSe$_3$  & 0.807 & $1.520 \pm 0.214$ & 1.281 & 0.723 & $0.934 \pm 0.135$ \\
EuDyS$_3$   & 0.838 & $1.872 \pm 0.319$ & 1.076 & 0.609 & $0.932 \pm 0.118$ \\
\textbf{PrScS$_3$}   & \textbf{0.822} & \textbf{$1.893 \pm 0.511$} & \textbf{1.586} & \textbf{0.899} & \textbf{$0.927 \pm 0.180$} \\
BaPrS$_3$   & 0.820 & $2.278 \pm 0.298$ & 0.783 & 0.445 & $0.922 \pm 0.166$ \\
SmZrS$_3$   & 0.840 & $1.846 \pm 0.314$ & 0.778 & 0.418 & $0.915 \pm 0.182$ \\
SmLuS$_3$   & 0.755 & $1.875 \pm 0.232$ & 1.076 & 0.609 & $0.914 \pm 0.184$ \\
CdCeS$_3$   & 0.825 & $2.126 \pm 0.391$ & 0.792 & 0.442 & $0.901 \pm 0.208$ \\
BaCeS$_3$   & 0.799 & $2.067 \pm 0.518$ & 0.783 & 0.445 & $0.894 \pm 0.200$ \\
\textbf{SmScS$_3$}   & \textbf{0.817} & \textbf{$1.927 \pm 0.353$} & \textbf{1.586} & \textbf{0.899} & \textbf{$0.880 \pm 0.230$} \\
SmYbS$_3$   & 0.817 & $1.896 \pm 0.334$ & 1.076 & 0.609 & $0.827 \pm 0.261$ \\
DyInS$_3$   & 0.844 & $1.681 \pm 0.205$ & 1.083 & 0.609 & $0.816 \pm 0.251$ \\
SrPrS$_3$   & 0.833 & $2.352 \pm 0.385$ & 0.925 & 0.530 & $0.760 \pm 0.309$ \\
EuYbS$_3$   & 0.817 & $1.762 \pm 0.467$ & 1.076 & 0.609 & $0.748 \pm 0.276$ \\
SnAlSe$_3$  & 0.829 & $2.019 \pm 0.292$ & 1.187 & 0.670 & $0.717 \pm 0.326$ \\
CaCeS$_3$   & 0.825 & $2.041 \pm 0.508$ & 1.805 & 1.023 & $0.703 \pm 0.305$ \\
SrCeS$_3$   & 0.812 & $2.102 \pm 0.612$ & 0.925 & 0.530 & $0.665 \pm 0.324$ \\
EuYbSe$_3$  & 0.815 & $1.323 \pm 0.257$ & 1.281 & 0.723 & $0.607 \pm 0.371$ \\
CuHfS$_3$   & 0.811 & $1.886 \pm 0.613$ & 0.383 & 0.193 & $0.585 \pm 0.344$ \\
CuHfSe$_3$  & 0.822 & $1.780 \pm 0.596$ & 0.589 & 0.307 & $0.570 \pm 0.360$ \\
ZnTbS$_3$   & 0.835 & $2.479 \pm 0.295$ & 0.759 & 0.428 & $0.560 \pm 0.358$ \\
CuZrSe$_3$  & 0.822 & $1.651 \pm 0.588$ & 0.589 & 0.307 & $0.518 \pm 0.337$ \\

\bottomrule
\end{tabular}
\end{table*}

Although the 54 \textit{CrystaLLM} candidates appear promising, their experimental realization may prove challenging. To address this, a final screening step was introduced to evaluate their experimental feasibility. High-throughput materials screening workflows often rely on thermodynamic descriptors, such as formation energy or energy above the convex hull, which are typically derived from DFT calculations and are therefore computationally demanding~\cite{jang2020_synthesizability}. As an initial attempt, available thermodynamic data were collected from open databases including the Materials Project (MP)~\cite{materials_project}, following the approach of Chen \textit{et al.}~\cite{Chen2024}. However, for the majority of the predicted and previously unexplored compositions, thermodynamic data were unavailable, underscoring a common limitation in evaluating virtual materials.

Recent work by Jung \textit{et al.} has demonstrated that relative synthesizability likelihood can be effectively assessed using positive-unlabeled (PU) machine-learning approaches~\cite{jang2020_synthesizability, gu2022_perovskite_synth, kim2024_llms_inorganic}. PU learning trains on known synthesized compounds as positive examples while treating unlabeled materials as potentially negative. This is particularly useful for synthesizability, since we know which compounds have been synthesized but do not have confirmed data on compounds that cannot be made. Compared to purely thermodynamic models, PU-based approaches exhibit fewer false positives, reflecting the influence of kinetic, chemical, and experimental factors beyond thermodynamic stability~\cite{kim2025_explainable_synth}.

The most recent model proposed by Jung \textit{et al.} converts crystallographic information files (CIFs) into textual representations and employs a fine-tuned GPT-4o-mini model to predict a crystal-likeness score (CLS), which provides a data-driven proxy for the relative synthesizability. This model reported strong performance on general inorganic datasets~\cite{kim2025_explainable_synth}. However, for perovskite-specific materials, a graph convolutional neural network (GCNN) has been shown to outperform language-based approaches, likely due to the relatively simple and repetitive connectivity of the perovskite-type structure~\cite{gu2022_perovskite_synth}. Accordingly, the GCNN model developed by Gu \textit{et al.} was employed in this work.

The GCNN was pre-trained on experimental structures from the ICSD and virtual perovskite-type crystals from MP, encompassing a broad range of chemistries including oxides, halides, hydrides, nitrides, and chalcogenides. The model CLS output can be seen as a statistical synthesizability indicator, where higher values indicate a greater likelihood of successful synthesis; a threshold of CLS = 0.5 has been proposed to identify synthetically plausible perovskites~\cite{gu2022_perovskite_synth}. No additional fine-tuning was performed, as the model is explicitly designed to generalize across perovskite-type chemistries. Crystal structures of the candidate set generated by \textit{CrystaLLM} were therefore used directly as input to the GCNN.

Chalcogenide candidates were ranked by CLS, most of them with values exceeding 0.5, which indicates greater statistical similarity to known synthesized crystals. Importantly, this indicator is not used to decide whether a perovskite structure exists, but to rank how the generated structures resemble experimentally realized perovskites, providing a complementary signal to geometric and topological screening. Moreover, all experimentally reported chalcogenide perovskites exhibit CLS values above 0.88, providing additional validation of the model for this class of materials and suggesting an empirical threshold for high synthesizability likelihood. For example, well-known Ba-based sulfide perovskites with B-site cations such as Zr, U, and Hf display CLS values of $0.99 \pm 0.02$.

Applying this stricter criterion reduces the candidate pool to approximately 30 compounds exhibiting high similarity to known experimentally reported  materials. In terms of compositional trends, higher CLS values are predominantly observed for sulfides, Ba- and La-based A-site cations, and Sc- and Tm-based B-site cations, whereas lower CLS values are associated with selenides, Cu- and Ce-based A-site compositions, and Yb- and Tb-based B-site cations.

\begin{figure}[H]
  \centering
  \includegraphics[width=\columnwidth]{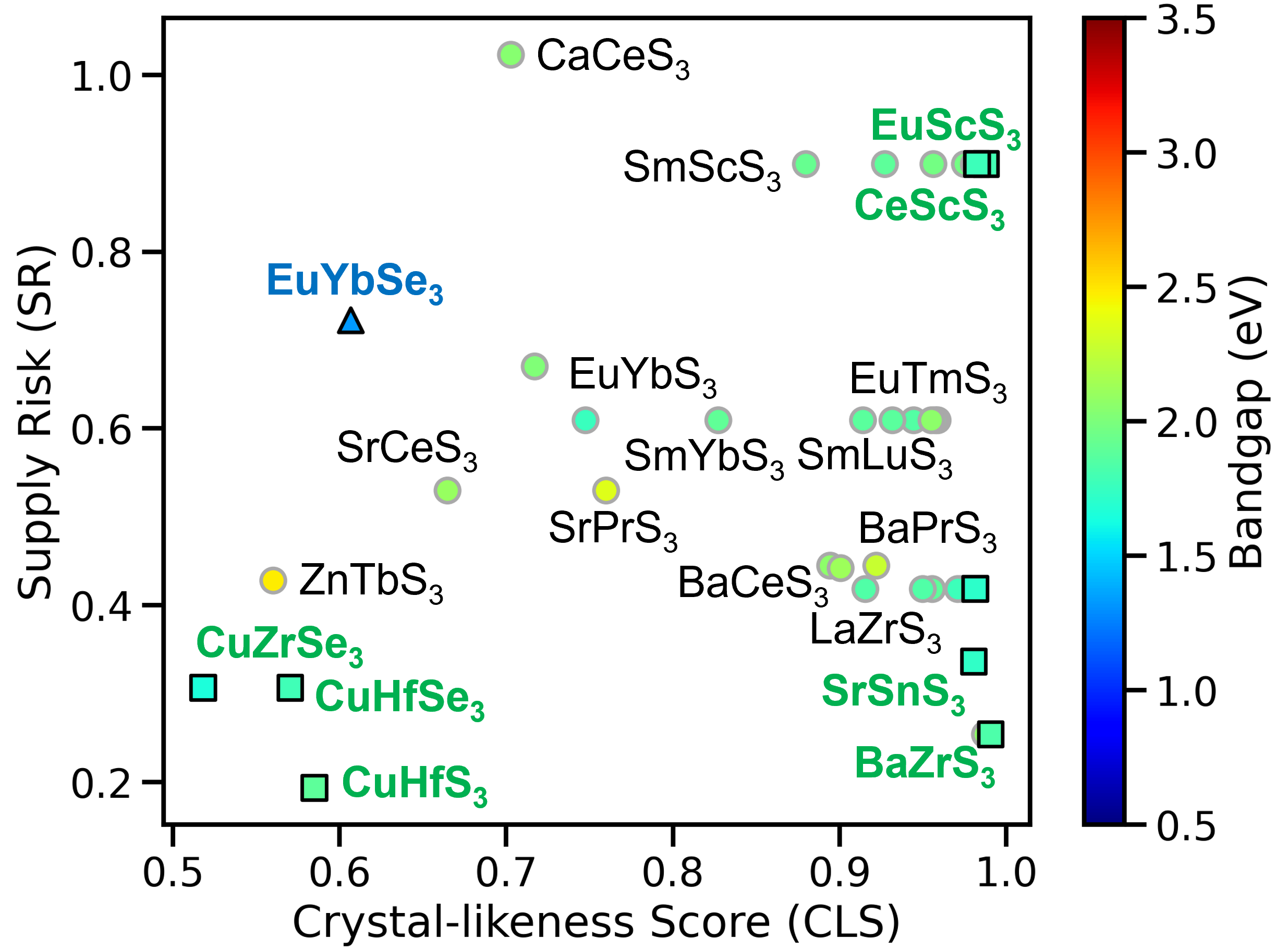}
  \caption{Supply risk (SR) as a function of the Crystal-likeness Score (CLS). The latter is a data-driven proxy for the relative synthesizability of the compound based on a pre-trained GCNN model \cite{gu2022_perovskite_synth}. The color code corresponds to the bandgap predicted by CrabNet \cite{wang2021_crabnet_explainable}. Squares and triangles indicate Pareto-optimal materials that simultaneously minimize SR, 1-CLS, and the bandgap deviation for tandem configurations ($E_g^{opt} = 1.71 eV$) and single junctions ($E_g^{opt} = 1.34 eV$), respectively.}
  \label{fig:sr_cls_eg_plot}
\end{figure}

Figure~\ref{fig:sr_cls_eg_plot} summarizes the combined assessment of the statistical synthesizability indicator, sustainability, and optoelectronic suitability for the most promising candidates. The figure reports CLS and SR values, with the predicted bandgap encoded by color and Pareto-optimal materials highlighted for single-junction (triangles) and tandem (squares) configurations. BaZrS$_3$ emerges as the most promising candidate for tandem applications. While this compound was included in the training dataset and thus cannot serve as an independent validation, its high ranking, together with the statistical analysis presented above, suggests that the individual models capture largely non-redundant aspects of material suitability. Furthermore, a Spearman rank correlation analysis over the 54 candidates confirms that pairwise correlations among the four screening metrics are uniformly weak ($|\rho| \leq 0.27$); only the $E_g$--CLS pair reaches marginal significance ($\rho = 0.27$, $p < 0.05$), while all remaining pairs are statistically non-significant (Fig.~S\ref{fig:corr_matrix}). This near-orthogonality indicates that each stage of the pipeline contributes independent discriminating information, reducing the likelihood that the ranking is governed by a latent common variable. Instead geometric stability, statistical synthesizability, sustainability, and optoelectronic suitability appear to capture fundamentally different chemical drivers rather than recapitulating the same compositional trend.

Table~\ref{tab:all_comp} summarizes the candidates along with their predicted bandgaps, CLS values, and estimated HHI and supply risks. These candidates should be interpreted as prioritized targets for experimental validation rather than definitive predictions of optoelectronic properties or thermodynamic stability. For instance, several yet-to-be-synthesized interesting compositions include EuScS$_3$, LaTbS$_3$, EuYbS$_3$, EuYbSe$_3$, CuHfS$_3$, CuHfSe$_3$, CuZrSe$_3$ (ranked by CLS).  Nonetheless, we would like to acknowledge that from a chemical perspective, some of these compositions may face additional challenges related to oxidation-state stability and charge balance, as  \texttt{pymatgen} oxidation-state assignments may not fully reflect redox thermodynamics. For instance, Cu is frequently stabilized in the +1 oxidation state (soft cation) in sulfides and selenides (soft bases), rendering Cu$^{2+}$-based chalcogenide perovskites potentially susceptible to reduction or mixed-valence behavior based on the hard and soft acids and bases (HSAB) theory \cite{miessler2014inorganic}. Similarly, Ce$^{4+}$ is strongly stabilized in oxides such as CeO$_2$ but is rarely observed in sulfide environments, where Ce$^{3+}$ phases dominate \cite{greenwood1997chemistry}. While the present screening workflow does not explicitly model redox stability, these considerations highlight the importance of chemical feasibility constraints and underscore the need for experimental verification of the most promising candidates.

Beyond the specific materials identified here, the proposed workflow illustrates a general and transferable strategy for data-driven materials screening in chemically complex and experimentally underexplored spaces. By combining interpretable analytical descriptors derived from experimental data, generative crystal structure prediction, composition-based property estimation, and experimental plausibility assessment, the pipeline enables systematic navigation of large compositional spaces while minimizing reliance on computationally intensive first-principles calculations. Importantly, each stage of the workflow produces physically interpretable outputs, allowing chemical intuition to be retained alongside machine-learning predictions.

The integration of sustainability metrics further extends the scope of the screening beyond traditional performance-driven criteria. These results highlight the importance of incorporating sustainability metrics at early stages of materials screening to better assess long-term technological viability, particularly for energy applications where material availability and supply-chain considerations can be as limiting as intrinsic performance. More broadly, this multi-objective framework illustrates how experimental knowledge, machine-learning models, and external socio-economic datasets can be combined to guide the identification of viable functional materials prior to experimental synthesis.

\section{Conclusions}

We present a data-driven framework to prioritize chalcogenide perovskite candidates by combining an analytically derived tolerance factor, generative crystal-structure prediction, composition-based bandgap estimation, and sustainability-oriented metrics. Rather than relying on a single screening criterion, the proposed workflow distributes decision-making across partially independent hypotheses, enabling structured reduction of chemical space while retaining physical interpretability at each stage.

The SISSO-derived tolerance factor $\tau^*$ significantly reduces false positives compared to classical geometrical descriptors while preserving recall for experimentally reported perovskites. Its intrinsic asymmetry between the A- and B-site radii provides chemically meaningful insight into the geometric constraints governing chalcogenide perovskites formation. Subsequent structure generation with \textit{CrystaLLM} resolves topological ambiguities within the tolerance-factor-stable region, refining the candidate pool to compounds capable of forming true corner-sharing $BX_6$ networks. Finally, integration of bandgap estimation, supply risk metrics, and crystal-likeness scoring enables multi-objective ranking tailored to photovoltaic applications.

The resulting candidates should be interpreted as prioritized targets for experimental validation rather than definitive predictions of thermodynamic stability or optoelectronic performance. In particular, the limited representation of selenide perovskites in the experimental dataset introduces additional uncertainty for ABSe$_3$ predictions. Nevertheless, the convergence of independent screening criteria and the weak pairwise correlations among them indicate that each stage contributes complementary information rather than reiterating the same chemical trends.

Beyond the specific compositions identified, this study illustrates how experimentally grounded descriptors, generative models, and socio-economic indicators can be integrated into a transparent and transferable screening strategy for chemically constrained materials spaces. Such approaches may help bridge the gap between theoretical prediction and experimental realization, particularly in underexplored material families where data scarcity and competing phases complicate discovery.

\section{Methods}

\subsection*{Data acquisition}

Experimental data were compiled from multiple literature and database sources to construct datasets for structural stability, electronic properties, experimental feasibility, and sustainability analysis (See the documentation in \nameref{sec:data}). For the determination of the tolerance factor $\tau^*$, the materials dataset consisted of experimentally reported $ABX_3$ compounds with X anions being halide or chalcogenides. Halide perovskite data were taken from Bartel \textit{et al.}~\cite{Bartel19}, while experimentally reported chalcogenide perovskites were collected from the work of Sopiha \textit{et al.}~\cite{Sopiha_2022}. In addition to compounds adopting the perovskite-type structure, experimentally reported $ABX_3$ compositions crystallizing in non-perovskite structure types were included as negative examples, enabling  discrimination between phases within the same stoichiometry.  Only oxygen-free compounds were considered throughout this study in order to focus on non-oxide perovskite-type systems. Halide compounds were included to increase the dataset size, while accepting the risk of bias toward ionic bonding characteristics.

Ionic radii were obtained from a human-curated dataset reported by Turnley \textit{et al.}, based on Shannon radii with extensions tailored to metal sulfides~\cite{turnley2024}. This dataset includes extrapolated values for missing entries and provides improved descriptions of metal-chalcogen bond distances, which are not adequately captured by conventional oxide-based radii tables. The use of this dataset reflects the increased covalent character of metal-chalcogen bonds relative to oxide and halide analogues. 

Experimental bandgap data were collected from three complementary sources: hybrid organic-inorganic halide perovskites from the NOMAD API~\cite{scheidgen_nomad_2023, marquez_perovskite_2024}, chalcogenide perovskites from the dataset of Sopiha \textit{et al.}~\cite{Sopiha_2022}, and a large set of chalcogenide semiconductors extracted using ChemDataExtractor~\cite{Dong2022}. Combining these sources enabled the construction of a broad, experimentally grounded dataset spanning both perovskite-type and chalcogenide materials, which is particularly valuable given the limited availability of chalcogenide perovskite data. 

Sustainability indicators were derived from publicly available datasets describing global commodity production and socio-environmental risk. Commodity production shares were taken from the US Geological Survey 2025 database~\cite{usgs2025}, environmental, social, and governance (ESG) indicators from the World Bank~\cite{worldbank2022_esgdata}. These datasets were combined to provide comparative estimates of material supply risk.

\subsection*{Tolerance factor determination}

A total of 283 experimentally reported $ABX_3$ compositions were used to develop and evaluate the tolerance factor $\tau^*$. The dataset was divided into an 80\% training set and a 20\% test set, with stratification to preserve similar distributions of anion types and experimentally stable perovskite-type phases across both subsets. Chalcogenide perovskites constitute approximately 21\% and 5\% of the training and test datasets for sulfides and selenides, respectively, corresponding to 27 experimentally reported compounds (26 ABS$_3$ and one ABSe$_3$). As a result, predictions for selenide perovskites are expected to carry a large intrinsic uncertainty.

Primary features were derived from elemental properties of the $A$, $B$, and $X$ species, including atomic radii ($r_A$, $r_B$, $r_X$), electronegativity ($\chi_A$, $\chi_B$, $\chi_X$), electron affinity ($EA_A$, $EA_B$, $EA_X$), ionization potential ($IP_A$, $IP_B$, $IP_X$), nuclear charge ($Z_A$, $Z_B$, $Z_X$), and oxidation state ($n_A$, $n_B$, $n_X$). In addition, computed energies of the highest occupied and lowest unoccupied atomic orbitals ($E^{\mathrm{HOMO}}$ and $E^{\mathrm{LUMO}}$) and orbital-resolved radii corresponding to the maxima of the radial distribution functions of $s$ and $p$ orbitals were included. Consistent with established perovskite screening approaches, additional derived features based on ratios of atomic radii and electronegativity differences were constructed~\cite{Jess2022, Bartel19}.

An initial feature selection was performed using a random forest classifier with Gini importance as the ranking criterion~\cite{Jiang_2025-vd}. Given the limited dataset size, the train-test split was randomized, and a grid search combined with cross-validation was used to optimize model hyperparameters with respect to the F$_1$-score. This procedure was repeated multiple times to assess robustness, and features were ranked according to their average normalized importance. The five most predictive features identified through this process were:
\begin{equation*}
    \Phi = \left[ \log\left(\frac{r_A}{r_B}\right), \frac{\Delta\chi_{B-X}}{\Delta\chi_{B-O}}, \frac{r_B}{r_X}, \frac{r_A}{r_X}, EA_A \right].
\end{equation*}

These features were subsequently used as inputs to the sure independence screening and sparsifying operator (SISSO) algorithm \cite{ouyang2018_sisso}. The implementation followed the approach of Bartel \textit{et al.}, employing an operator space consisting of 11 mathematical operations: $[+, -, |-|, \times, \div, \exp, \square^{-1}, \square^2, \square^3, \sqrt{\square}, \log]$~\cite{Bartel19}. Four iterations of sure independence screening were performed, retaining 200 residuals per iteration, yielding a pool of approximately 3000 candidate descriptors.

A decision tree classifier with a maximum depth of one was employed as the sparsifying operator to select a single optimal descriptor from the candidate pool. This shallow-tree approach was chosen to preserve descriptor interpretability and reduce sensitivity to individual data points, particularly given the limited and heterogeneous nature of the experimental dataset. Classifiers were trained using the entropy criterion, with class weights adjusted according to class frequency to account for the underrepresentation of chalcogenide perovskites. This strategy acknowledges that experimental classification of perovskite versus non-perovskite is often ambiguous due to variations in synthesis conditions.

Finally, a compositional screening of candidate materials was performed by enumerating 1392 charge-balanced $ABX_3$ compounds formed from combinations of metals and chalcogenides satisfying the perovskite-type stoichiometry~\cite{Schilling_Wilhelmi_2025}. Oxidation states were assigned using the prediction scheme implemented in \texttt{pymatgen}, resulting in a final list of 181 compounds predicted to be structurally stable according to the derived $\tau^*$ criterion.

\subsection*{Crystal structure generation}

The crystal structures of the 181 compounds predicted to be stable based on the $\tau^*$ criterion were generated from their chemical formulae using \textit{CrystaLLM}~\cite{antunes_crystal_2024}. The large pre-trained model was employed, and four crystallographic unit cells were generated for each composition \cite{Antunesetal2024}. CrystaLLM is based on an autoregressive large language model trained to generate Crystallographic Information Files (CIFs) and has been trained on over three million compounds drawn from open-access density functional theory databases, including the Materials Project \cite{materials_project}, NOMAD\cite{scheidgen_nomad_2023}, and OQMD~\cite{saal2013_oqmd, antunes_crystal_2024}.

From the generated structures, 116 compounds were predicted to adopt the orthorhombic \textit{Pnma} space group, which is commonly associated with distorted perovskite-type structures. However, within this space group, the BX$_6$ octahedra may be arranged in either edge-sharing or corner-sharing configurations, with only the latter corresponding to the true perovskite-type structure. To distinguish between these motifs, the connectivity of the octahedral network was analyzed for each predicted structure. Only structures exhibiting a corner-sharing BX$_6$ framework were retained, resulting in a final set of 54 compounds with a generated perovskite-type crystal structure.

It is important to note that this classification relies on generative structural predictions rather than experimentally relaxed geometries, and therefore introduces an additional source of uncertainty. In particular, subtle distortions and alternative low-energy phases may not be fully captured at this stage. Consequently, the predicted crystal structures are intended to serve as screening-level indicators of perovskite-type structure  rather than definitive structural assignments.

\subsection*{Bandgap estimation}
Bandgaps were estimated using a composition-based machine-learning model trained on experimentally reported data. The initial dataset comprised 50\,662 reported materials, including hybrid organic-inorganic halide perovskites, chalcogenide perovskites, and chalcogenide semiconductors~\cite{scheidgen_nomad_2023, Sopiha_2022, Dong2022, Shabihetal2026}. The latter ensures diverse data that encourages generalizable composition–property relationships rather than narrow domain fitting. Materials with reported bandgaps below 1~eV were excluded to focus on semiconductors relevant for photovoltaic applications. In addition, for each composition, reported values more than 0.3~eV below the median bandgap were discarded to suppress anomalously low values likely related to DFT-derived quantities. Compounds were subsequently  grouped by chemical formula, resulting in a curated dataset of 3\,628 unique compositions, each assigned a representative bandgap given by the median of the reported values. This dataset was randomly divided into training, validation, and test subsets.

A CrabNet model~\cite{Wang2021crabnet} was trained using several compositional encoding schemes including magpie, mat2vec, onehot, random\_200 and Pettifor descriptors~\cite{Tshitoyan2019, wang2021_crabnet_explainable, Cerqueira2025, ward2016_magpie, tshitoyan2019_mat2vec}. The model architecture consisted of three attention layers with an embedding dimension of 512 and four attention heads, resulting in approximately 12 million trainable parameters. Training was performed on a CPU using a batch size of 128 and a cyclical learning rate schedule, with early stopping enabled based on validation mean absolute error (MAE). Finally, model performance was evaluated on the held-out test set, yielding a mean absolute error (MAE) of 248~meV for a combined embedding representation with mat2vec and Pettifor. This combination was chosen as training solely with mat2vec embedding resulted in a MAE of 256~meV. For comparison, other models were trained only with the halide perovskites or chalcogenide semiconductors datasets and baseline models predicting the median or mean bandgap for all compositions were assessed. Table~\ref{tab:MAE_models} shows the respective test-set MAE and R$^2$ for these models.

As conventional random train/test split for this case yields zero or one chalcogenide perovskite in the test set, any error estimate on the target chemical space is statistically unreliable. Thus to obtain a more rigorous estimate of the model's ability to extrapolate to unseen chalcogenide perovskites, we employ Leave-One-Out Cross-Validation (LOOCV): in each of the nine folds, one chalcogenide perovskite composition is withheld entirely from training, and CrabNet is retrained from scratch on the remaining corpus of $\sim$3\,600 compounds (halide perovskites and chalcogenide semiconductors), then used to predict the bandgap of the held-out compound. The median experimental bandgap per formula, together with its inter-measurement standard deviation as a proxy for experimental uncertainty, is used as the reference value. This procedure yields a LOOCV MAE of 281~meV across the nine compositions, providing a more reliable estimate of the model's performance on the target class of materials.

Given the compositional nature of the model and the structural diversity of the training data, the predicted bandgaps are intended as approximate screening-level indicators rather than quantitative optoelectronic predictions. Nevertheless, this approach enables rapid identification of candidate materials with bandgaps compatible with photovoltaic applications within the broader screening workflow.

\begin{table}[H]
\centering
\caption{Performance comparison of composition-based bandgap models and trivial baselines.
Models were trained on different subsets of the experimental dataset.
All metrics are reported for the held-out test set. MAE corresponds to the Mean absolute error, while R$^2$ is the coefficient of determination.}
\label{tab:MAE_models}
\begin{tabular}{lccc}
\toprule
Model & $n_{\mathrm{train}}$ & MAE (meV) & $R^2$ \\
\midrule
Full dataset & 3628 & 248 & 0.423 \\
Halide-only & 1263 & 103 & 0.406 \\
Chalcogenides-only & 1630 & 430 & 0.290 \\
Median baseline & -- & 443 & -- \\
Mean baseline & -- & 472 & -- \\
\bottomrule
\end{tabular}
\end{table}

\subsection*{Sustainability analysis}

To assess the sustainability of the screened chalcogenide perovskites, supply risk indicators were calculated following the methodology proposed by Nominé \textit{et al.}~\cite{nomine2023}. Two complementary metrics were evaluated using publicly available datasets describing global commodity production and socio-environmental risk.

The Herfindahl-Hirschman Index (HHI) was used to quantify market concentration for each elemental commodity and was calculated as the sum of the squared country-level production shares. Production data were taken from the most recent US Geological Survey dataset (USGS 2025), corresponding to 2023 global production~\cite{usgs2025}. Commodity-level HHI values were subsequently mapped to elemental species within each candidate compound.

Environmental, social, and governance (ESG) risk was estimated using 18 normalized country-level indicators obtained from the World Bank~\cite{worldbank2022_esgdata}, grouped into environmental, social, and governance categories with six indicators per category. These indicators serve as proxies for ESG conditions associated with raw material production and reflect aggregate national-level trends rather than site-specific mining practices. The selected indicators are listed below.

\begin{itemize}
    \item \textbf{Social score:} average of indicators including internet usage, labor force participation rate, life expectancy at birth, proportion of seats held by women in national parliaments, school enrollment, and unemployment.
    \item \textbf{Environmental score:} average of indicators including value added by agriculture, forestry, and fishing; electricity production from coal; net energy imports; per capita energy use; fossil fuel energy consumption; and extent of terrestrial and marine protected areas.
    \item \textbf{Governance score:} average of World Bank governance estimates for control of corruption, government effectiveness, political stability and absence of violence or terrorism, regulatory quality, rule of law, and voice and accountability.
\end{itemize}

The average ESG score was combined with country-level production shares as weights in the HHI formulation to estimate an overall supply risk for each elemental commodity. For a given compound, the total supply risk was calculated as the sum of the elemental supply risks. These sustainability metrics provide a comparative framework for prioritizing candidate materials rather than absolute assessments of environmental impact.

\subsection{Experimental plausibility assessment}

Experimental plausibility was evaluated using a machine-learning approach based on positive–unlabeled (PU) learning, which has been shown to effectively capture experimental realizability by learning from known synthesized materials without requiring explicit negative examples. In this work, we employed a pre-trained graph convolutional neural network (GCNN) model developed by Gu \textit{et al.}~\cite{gu2022_perovskite_synth}, specifically designed for perovskite-type crystal structures.
GCNNs are particularly well suited for this task, as they encode local coordination environments and connectivity patterns that are central to perovskite-type chemistry, allowing structurally meaningful comparison across diverse compositions.

The model outputs a crystal-likeness score (CLS), which quantifies the likelihood that a given crystal structure resembles experimentally observed perovskite-type compounds. Higher CLS values correspond to greater experimental feasibility.
Importantly, CLS is not a thermodynamic quantity and does not directly account for kinetic barriers, synthesis pathways, or processing conditions; rather, it serves as a comparative metric to prioritize candidate materials for experimental investigation.

Crystal structures generated by \textit{CrystaLLM} were used directly as input to the GCNN.
No geometry optimization or energy minimization was performed prior to evaluation, as the purpose of the synthesizability assessment is to analyze coordination features rather than energetic stability. No additional fine-tuning of the model was performed. A CLS value of 0.5 has previously been proposed as a threshold indicating synthetically plausible perovskite-type structures~\cite{gu2022_perovskite_synth}; however, in this work CLS values are interpreted comparatively rather than as strict pass–fail criteria. Validation against experimentally reported chalcogenide perovskites indicates that known compounds consistently exhibit CLS values above 0.88, providing an empirical reference for high synthesizability likelihood within this class of materials.
Accordingly, CLS was used as a ranking and filtering metric to guide candidate prioritization.

\section*{Author contributions}
DAG: data curation, formal analysis, investigation, methodology, software, visualization and writing - original draft. LH:   writing - review \& editing. LA: supervision, writing - review \& editing. SaS: supervision, writing - review \& editing. JAM: conceptualization, software, visualization, methodology, supervision, writing - review \& editing.

\section*{Conflicts of interest}
There are no conflicts to declare.

\section{Data \& Code Availability} 
\phantomsection
\label{sec:data}

The curated datasets and code developed in this work are openly available at the GitHub repository~\cite{garzon2026chalc_screening}: \url{https://github.com/GarzonDiegoFEUP/chalcogenide-perovskite-screening}. An interactive version of the figures is available in the documentation at the GitHub repository. A citable archived version (v1.0.1) is deposited on Zenodo at \url{https://doi.org/10.5281/zenodo.18743650}. 

\section*{Acknowledgements}

DAG acknowledges the support by FCT - Fundação para a Ciência e Tecnologia, I.P. with the project reference 2023.00258.BD, and DOI identifier https://doi.org/10.54499/2023.00258.BD.  Authors acknowledge the COST Action Research and International Networking project "Emerging Inorganic Chalcogenides for Photovoltaics (RENEW-PV)," CA21148, supported by COST (European Cooperation in Science and Technology).

\printbibliography
\end{multicols}


\clearpage
\onecolumn

\section*{Supplementary Information}
\phantomsection
\addcontentsline{toc}{section}{Supplementary Information}

This Supplementary Information provides additional analyses supporting the main results of the screening workflow.
~\Cref{sec:S1} reports the classification performance of the tolerance-factor models and the logistic-calibrated stability probabilities. ~\Cref{sec:S2} presents structural validation of the \textit{CrystaLLM}-generated geometries against experimental ICSD data. ~\Cref{sec:S3} documents robustness tests and ablation studies for the bandgap model, including dataset-size dependence, encoder comparisons, parity analysis, PCA analysis, and null-model baselines.
Finally, ~\Cref{sec:S4} quantifies the complementarity of the screening criteria through rank-correlation analyses.

\renewcommand{\figurename}{Figure S}
\renewcommand{\tablename}{Table S}
\setcounter{figure}{0}
\setcounter{table}{0}

\clearpage
\SIsection{Support for tolerance-factor classification}{sec:S1}
This section provides additional validation of the tolerance-factor classifiers, including confusion matrices and stability-probability maps derived from $\tau^*$.
\phantomsection

\begin{figure}[!htbp]
  \centering
  \includegraphics[width=0.8\textwidth]{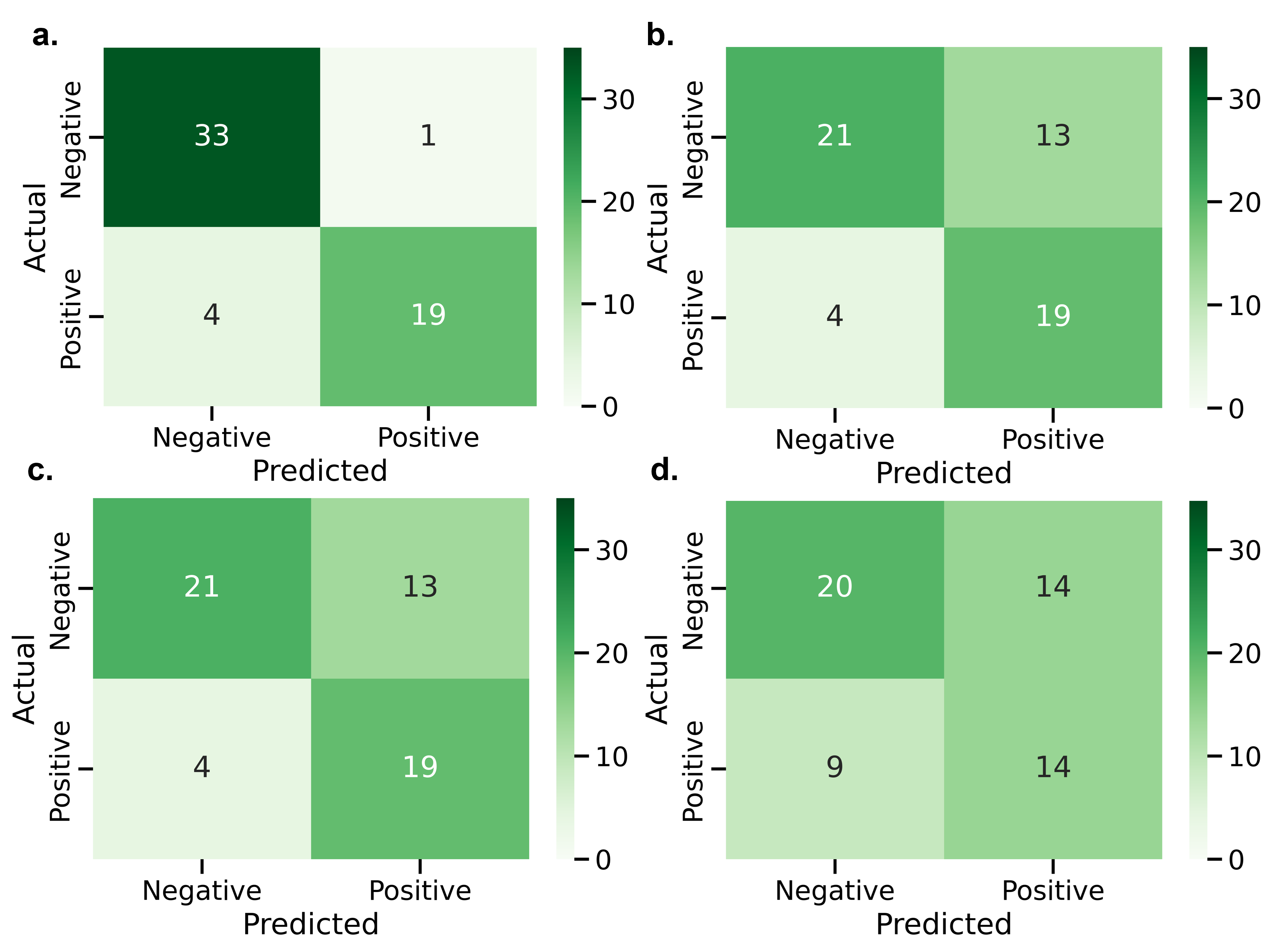}
  \caption{
  Confusion matrices for perovskite stability classification using four tolerance factors:
  (a) SISSO-derived $\tau^*$,
  (b) Bartel \textit{et al.} $t_{\text{Bartel}}$,
  (c) Jess \textit{et al.} $t_{\text{Jess}}$,
  and (d) Goldschmidt $t_{\text{Goldschmidt}}$.
  Matrices follow the convention $\left[\left[\mathrm{TN},\mathrm{FP}\right],\left[\mathrm{FN},\mathrm{TP}\right]\right]$,
  where TN is true negative, FP is false positive, FN is false negative, and TP is true positive.
  }
  \label{fig:cm_tolerance_factor}
\end{figure}

\begin{figure}[!htbp]
  \centering
  \includegraphics[width=\textwidth]{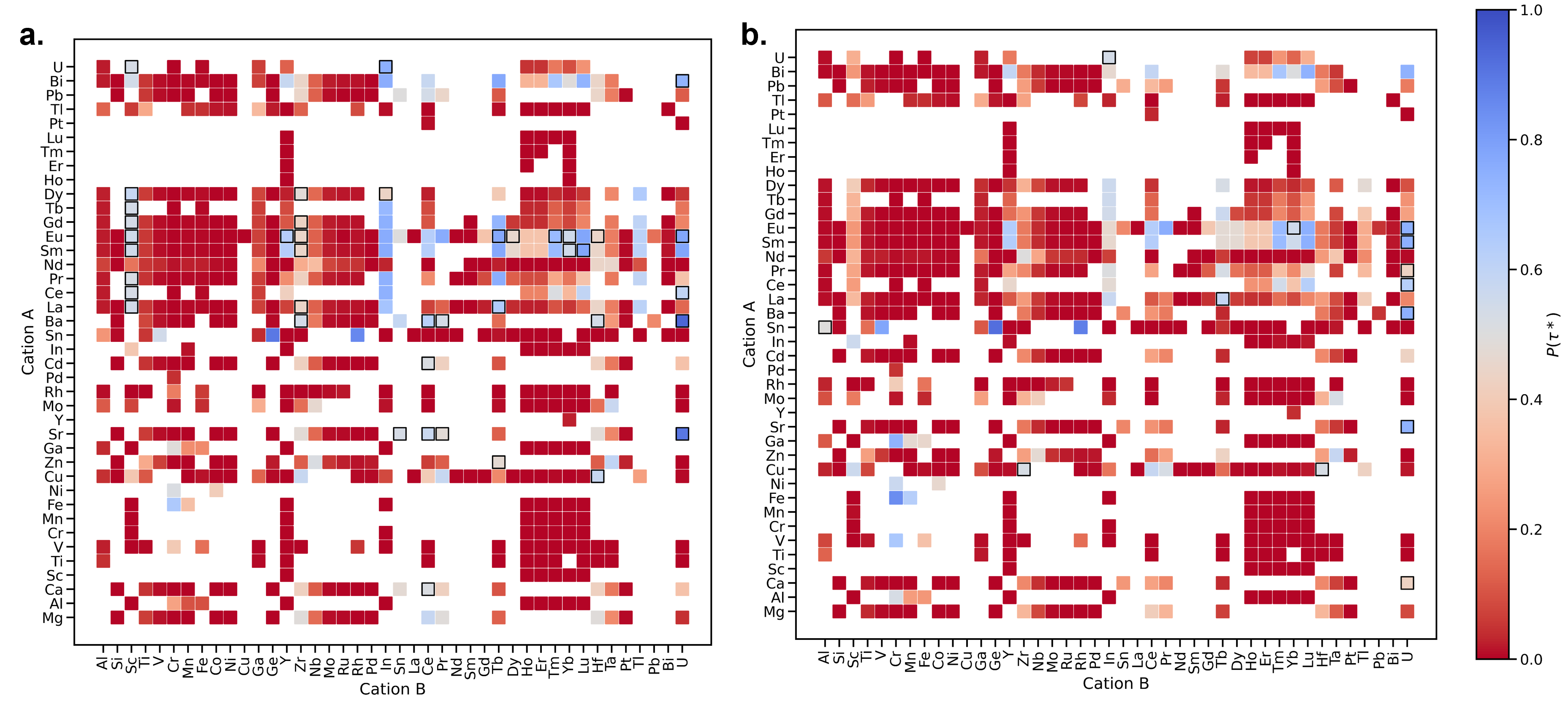}
  \caption{
  Logistic-calibrated probability of perovskite-type structural stability across the enumerated chemical space based on the SISSO-derived tolerance factor $\tau^*$:
  (a) ABS$_3$ and (b) ABSe$_3$ compositions.
  Squares outlined in black correspond to compositions for which \textit{CrystaLLM} generated a corner-sharing perovskite-type structure.
  }
  \label{fig:matrix_prob}
\end{figure}

\clearpage
\SIsection{Structural validation against experimental ICSD data}{sec:S2}
This section compares lattice parameters of \textit{CrystaLLM}-generated structures against ICSD-reported experimental values for known chalcogenide perovskites.
\phantomsection

\begin{figure}[!htbp]
  \centering
  \includegraphics[width=0.85\textwidth]{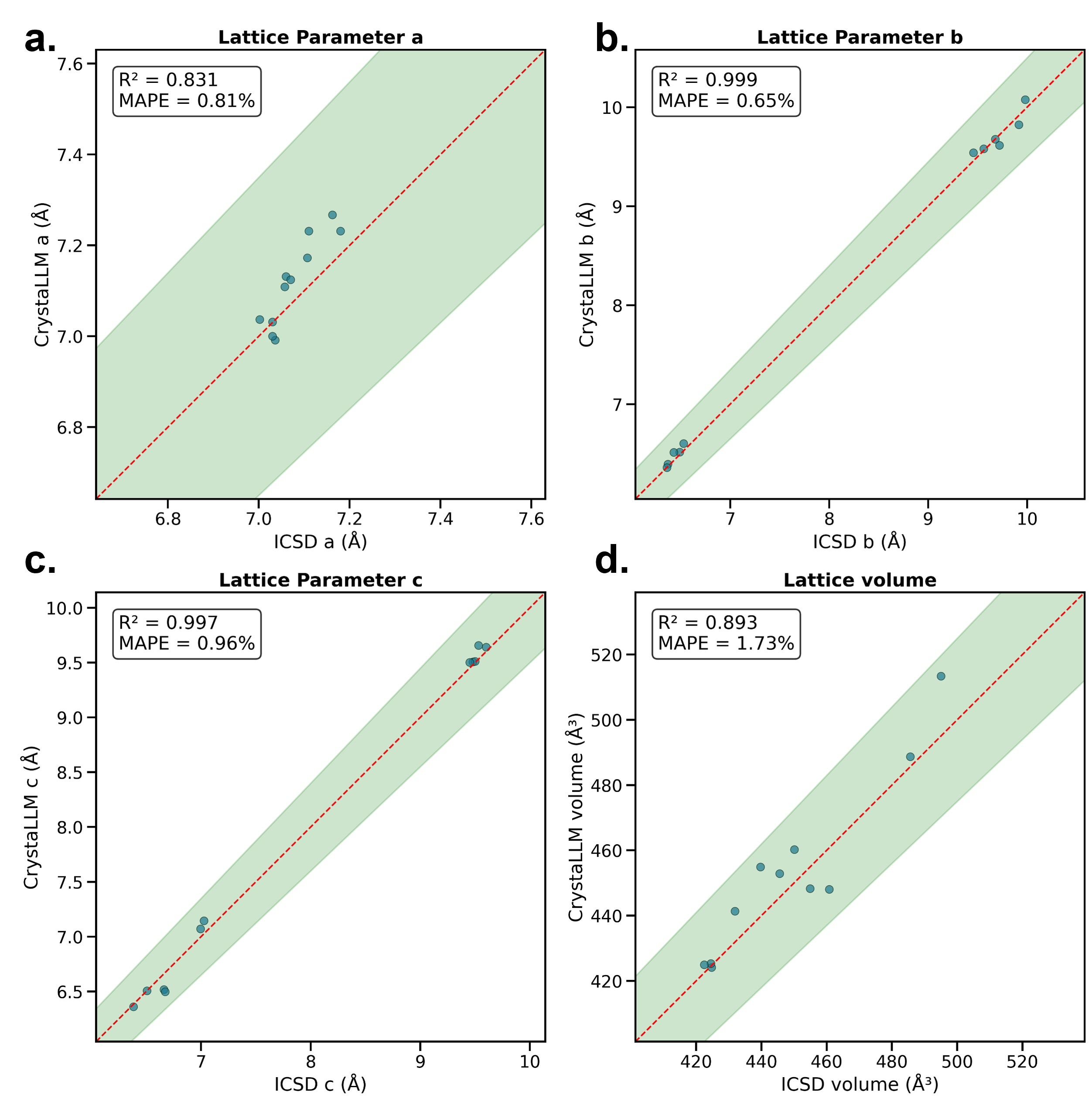}
  \caption{
  Parity plots comparing lattice parameters and unit-cell volume of \textit{CrystaLLM}-generated structures against ICSD-reported experimental data for experimentally observed chalcogenide perovskites:
  (a) $a$,
  (b) $b$,
  (c) $c$,
  and (d) unit-cell volume.
  Each panel reports $R^2$ and the mean absolute percentage error (MAPE).
  The green shaded region indicates a $\pm 5\%$ deviation from perfect agreement.
  }
  \label{fig:crystal_ICSD}
\end{figure}

\clearpage
\SIsection{Bandgap model validation and ablations}{sec:S3}
This section reports robustness tests for the bandgap model, including learning-curve behavior, encoder sensitivity, parity analysis, PCA, and baseline comparisons.
\phantomsection

\begin{figure}[!htbp]
  \centering
  \includegraphics[width=\textwidth]{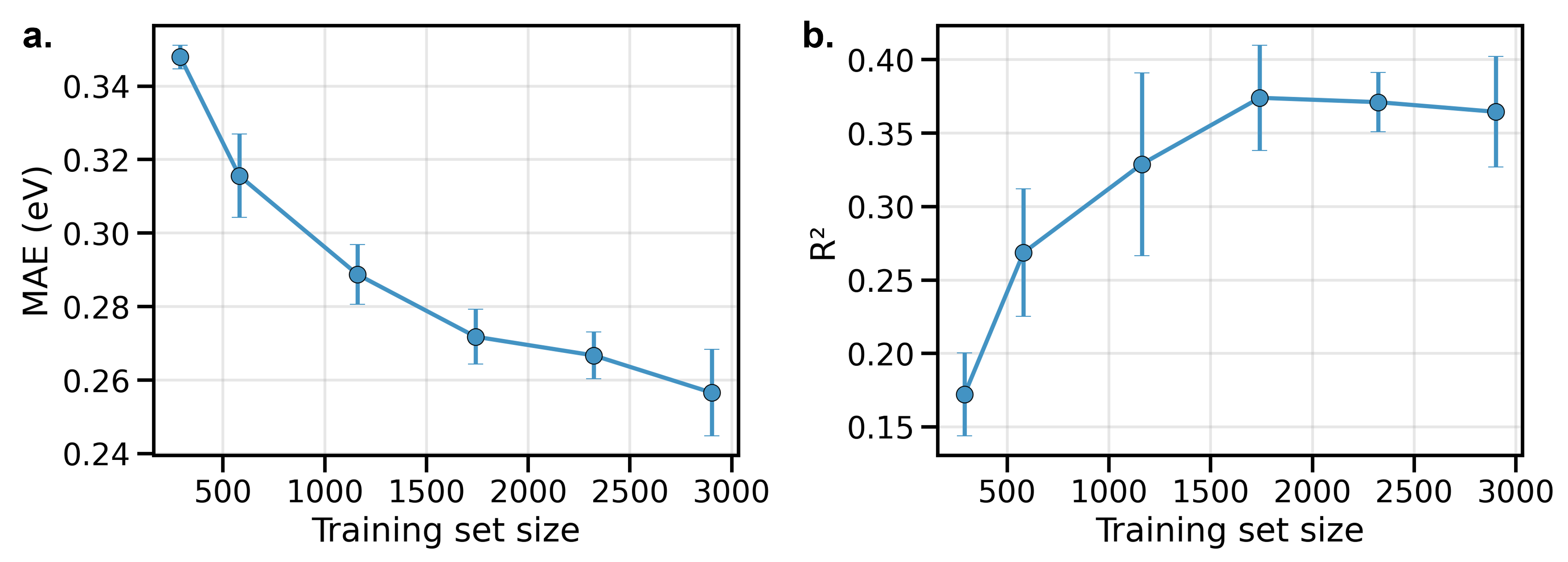}
  \caption{
  Influence of training dataset size on bandgap prediction performance.
  (a) Mean absolute error (MAE) on the held-out test set as a function of training set size.
  (b) Corresponding coefficient of determination ($R^2$).
  Error bars indicate standard deviation across repeated random splits.
  The results show gradual improvement with increasing dataset size and suggest convergence beyond approximately 2000 training samples.
  }
  \label{fig:dataset_size}
\end{figure}

\begin{figure}[!htbp]
  \centering
  \includegraphics[width=0.7\textwidth]{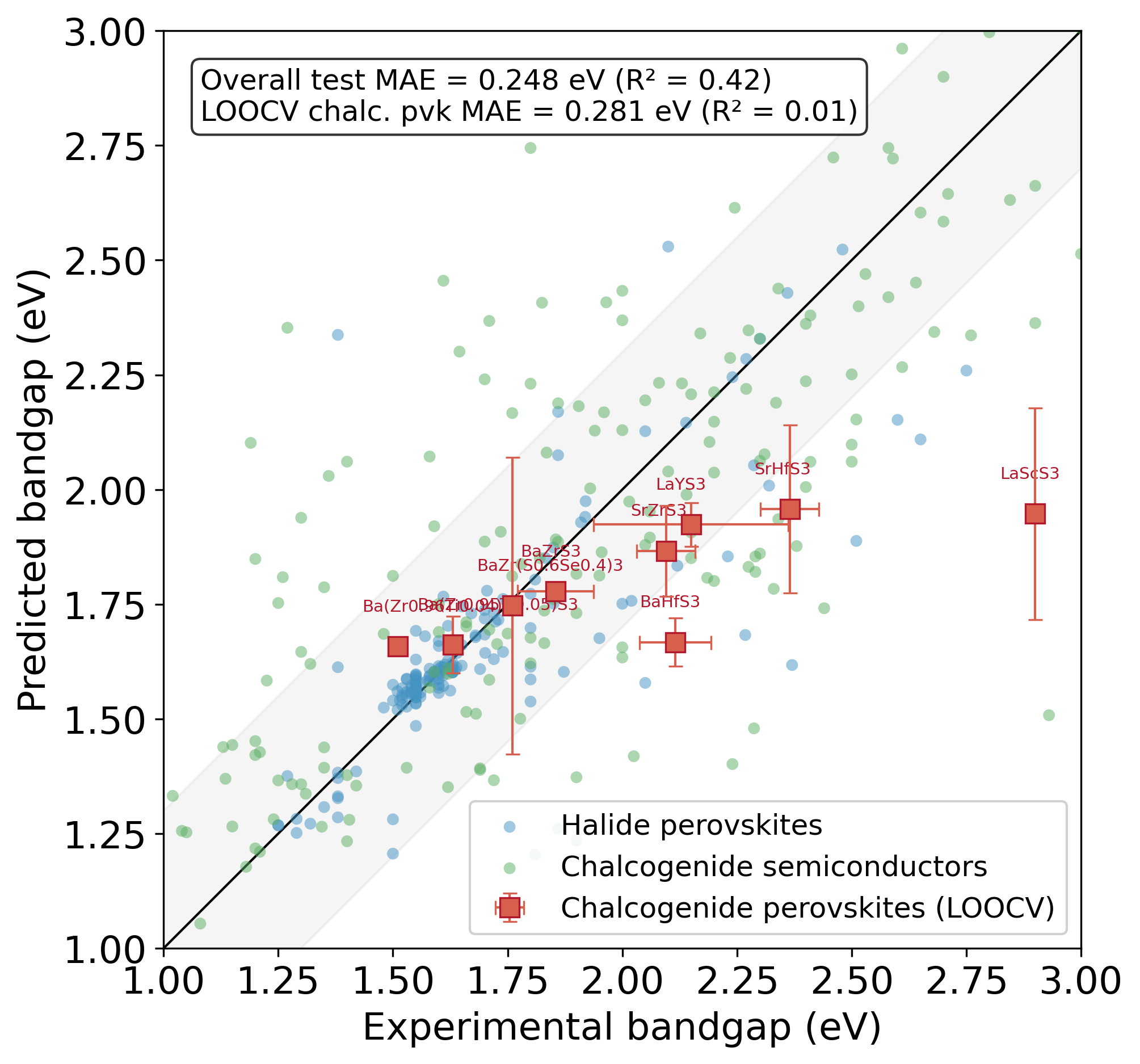}
  \caption{
  Parity plot for the bandgap prediction model.
  Predicted versus experimental bandgaps are shown for halide perovskites (blue), chalcogenide semiconductors (green), and experimentally reported chalcogenide perovskites (red squares; LOOCV).
  For chalcogenide perovskites, predictions were obtained using a leave-one-out cross-validation (LOOCV) procedure due to the limited dataset size; error bars indicate the standard deviation across LOOCV folds.
  The solid black line indicates perfect agreement, and the shaded region represents a $\pm 0.3$~eV deviation.
  The overall test-set MAE and $R^2$ are reported for the full held-out dataset, while separate metrics are shown for the LOOCV evaluation of chalcogenide perovskites.
  }
  \label{fig:bg_parity}
\end{figure}

\begin{table}[!htbp]
\centering
\caption{
Null-model baselines for bandgap prediction.
CrabNet performance is compared against trivial predictors that always output the mean or median bandgap of the corresponding evaluation subset.
Results are reported for (i) the held-out test set and (ii) chalcogenide perovskites evaluated using leave-one-out cross-validation (LOOCV).
}
\label{tab:null_baselines}
\begin{tabular}{l l c}
\toprule
Subset & Strategy & MAE (meV) \\
\midrule
Test set & Median predictor & 443 \\
Test set & Mean predictor & 472 \\
Test set & CrabNet & 248 \\
\midrule
Chalcogenide perovskites (LOOCV) & Median predictor & 308 \\
Chalcogenide perovskites (LOOCV) & Mean predictor & 314 \\
Chalcogenide perovskites (LOOCV) & CrabNet (LOOCV) & 281 \\
\bottomrule
\end{tabular}
\end{table}

\begin{figure}[!htbp]
  \centering
  \includegraphics[width=0.7\textwidth]{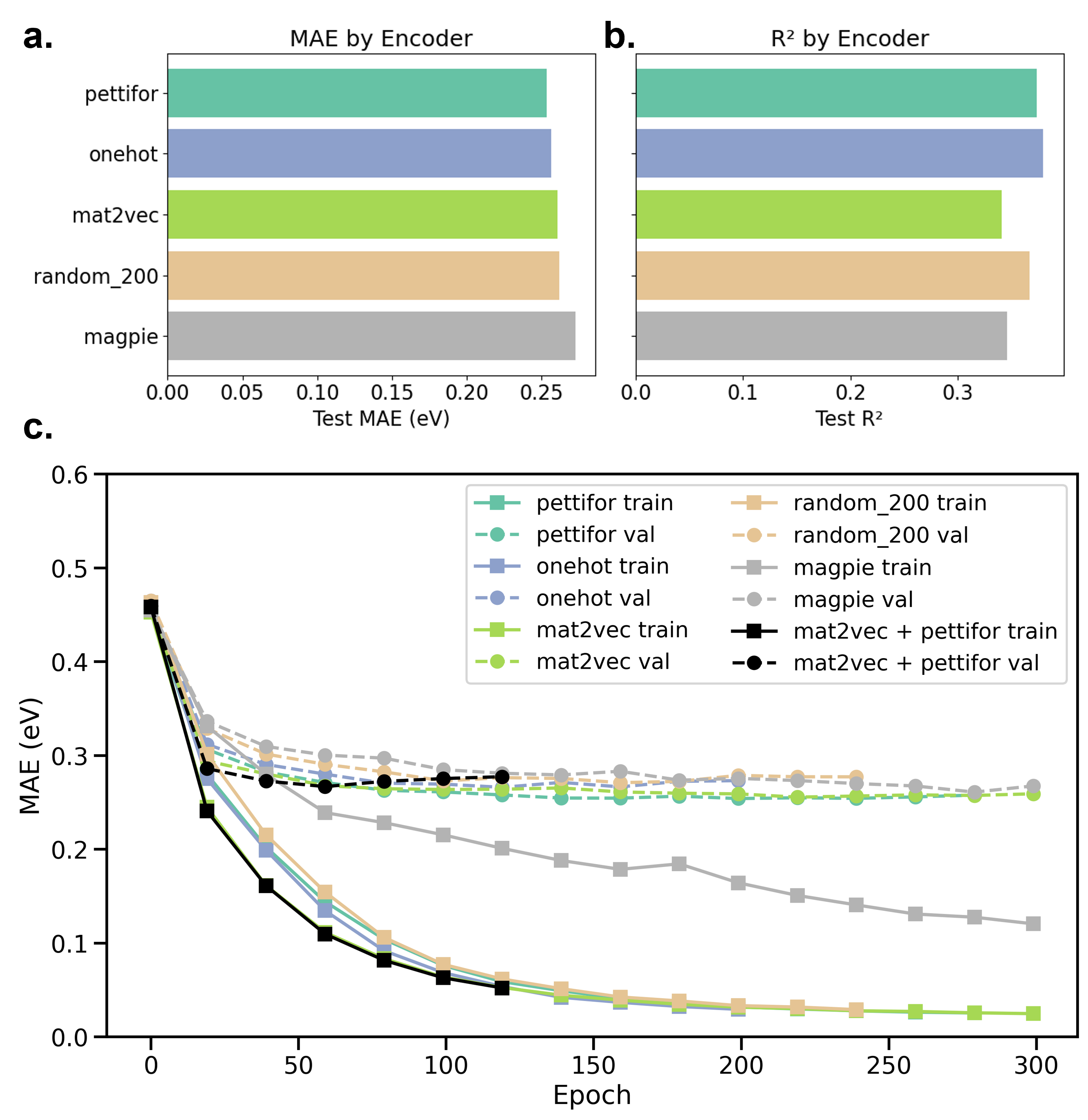}
  \caption{
  Comparison of element encoders used for CrabNet bandgap prediction.
  (a) Test-set mean absolute error (MAE) across different encoders.
  (b) Test-set coefficient of determination ($R^2$) for the same models.
  (c) Training and validation MAE as a function of epoch for selected encoders (solid: training; dashed: validation), including individual encoders and a combined mat2vec + Pettifor representation.
  All models converge to comparable validation performance, indicating limited sensitivity of final predictive accuracy to the choice of initial element encoding.
  }
  \label{fig:comp_encoders}
\end{figure}

\begin{figure}[!htbp]
  \centering
  \includegraphics[width=\textwidth]{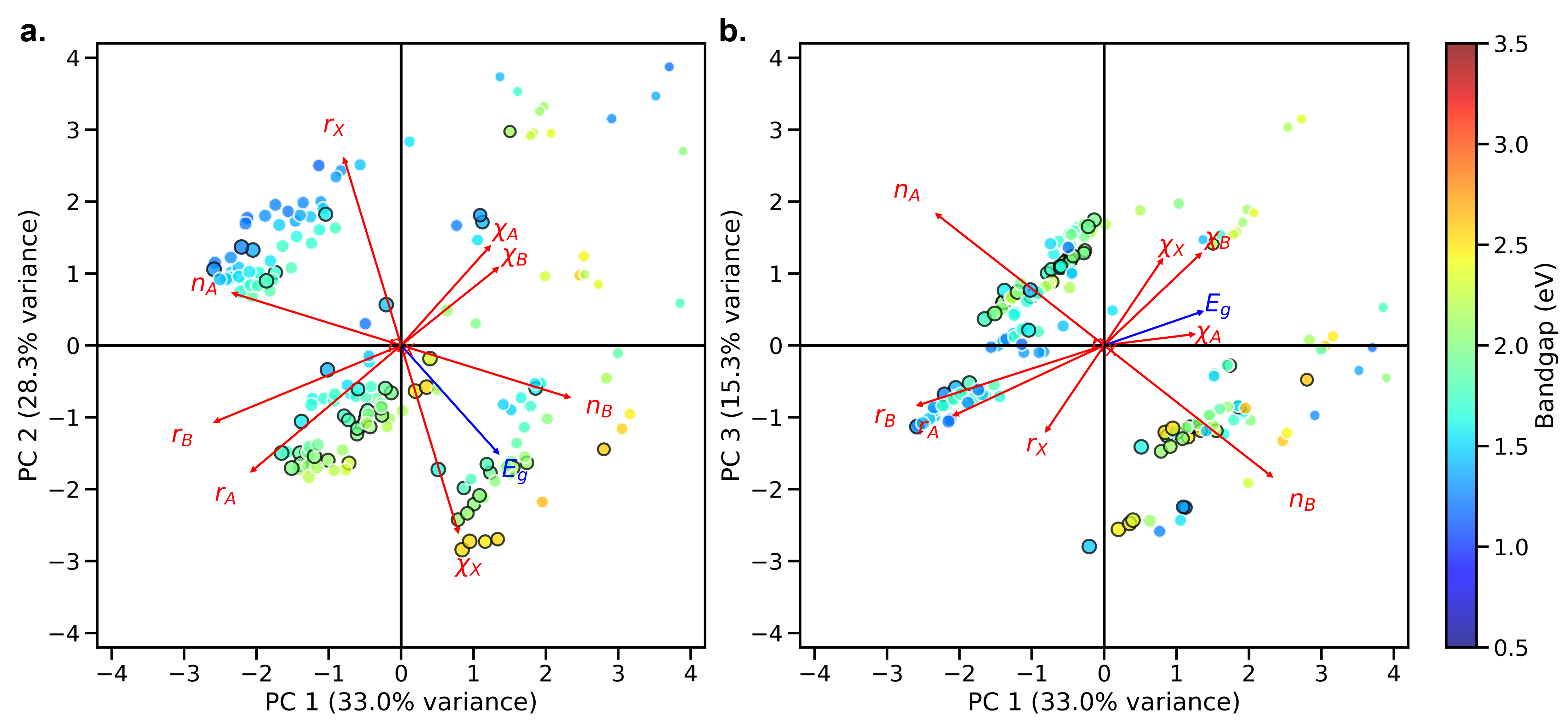}
  \caption{
  Principal component analysis (PCA) of the predicted bandgaps for the $\tau^*$-stable chalcogenide compositions.
  Color indicates the predicted bandgap value and marker size scales with the B-site ionic radius.
  Points outlined in black correspond to compositions for which \textit{CrystaLLM} generated a corner-sharing perovskite-type structure.
  }
  \label{fig:pca_analysis}
\end{figure}

\SIsection{Complementarity of screening metrics}{sec:S4}
This section assesses whether the screening criteria provide non-redundant information by computing pairwise Spearman rank correlations across the candidate set.

\begin{figure}[!htbp]
  \centering
  \includegraphics[width=\textwidth]{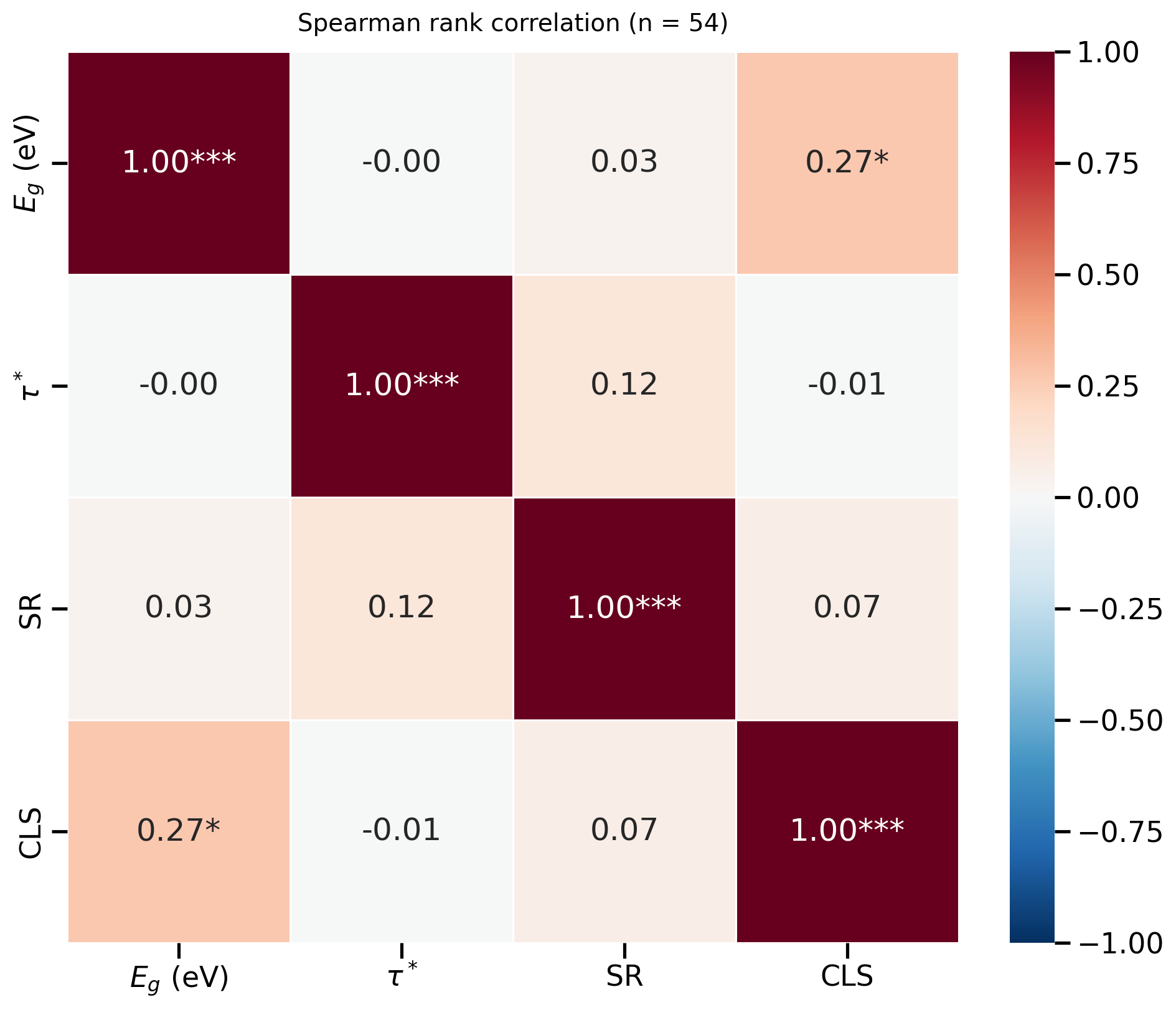}
  \caption{
  Spearman rank correlation matrix ($n = 54$) of screening metrics for the \textit{CrystaLLM}-predicted perovskite-type candidate set.
  Significance levels are indicated by asterisks ($^{*}p < 0.05$, $^{**}p < 0.01$, $^{***}p < 0.001$).
  $E_g$ denotes CrabNet-predicted bandgap, $\tau^*$ is the SISSO-derived tolerance factor, SR is the supply risk, and CLS is the crystal-likeness score from the pre-trained GCNN model.
  All off-diagonal correlations are weak ($|\rho| \leq 0.27$), indicating that the screening criteria capture largely complementary aspects of material suitability.
  }
  \label{fig:corr_matrix}
\end{figure}

\end{document}